\documentclass[a4paper, 12 pt]{article}
\usepackage[T1]{fontenc}
\usepackage[utf8]{inputenc}
\usepackage[english]{babel}
\usepackage{amsmath}
\usepackage{amssymb}
\usepackage{braket}
\usepackage{mathtools}
\usepackage{dsfont}
\usepackage{subcaption}
\usepackage{booktabs}
\usepackage{yfonts}
\usepackage{setspace}
\usepackage{cite}
\usepackage{verbatim}

\usepackage{amstext} 
\usepackage{array}   
\newcolumntype{C}{>{$}c<{$}} 

\usepackage{geometry}
\usepackage[usenames,dvipsnames]{color}
\usepackage{hyperref}
\hypersetup{
  colorlinks,
  citecolor=Blue,
  linkcolor=Blue,
  urlcolor=Blue}

\numberwithin{equation}{section}

\def\be{\begin{equation}}
\def\ee{\end{equation}}

\newcommand{\diff}{\mathrm{d}}

\definecolor{darkviolet}{rgb}{0.58, 0.0, 0.83}

\def\kk{{\mathsf{k}}}

\begin{document}

\pagestyle{empty}

\begin{center}

$\,$
\vskip 1.5cm

{\LARGE{\bf Three-charge black holes from the worldsheet}}

\vskip 1cm

{\large
Stefano Massai,${}^{\textrm a,b}$
Enrico Turetta${}^{\textrm a,b}$
}

\vskip 1cm

\end{center}

\renewcommand{\thefootnote}{\arabic{footnote}}

\begin{center}
$^{\textrm a}${\it Dipartimento di Fisica e Astronomia ``Galileo Galilei'', Universit\`a di Padova,\\Via Marzolo 8, 35131 Padova, Italy}\,,\\[2mm]
$^{\textrm b}$ {\it INFN, Sezione di Padova, Via Marzolo 8, 35131 Padova, Italy}\,.

\vskip 3cm

 {\bf Abstract} 
\end{center}

{\noindent 
We consider the general asymptotically flat, doubly-rotating, three-charge black hole uplifted to a family of NS5-F1-P black brane solutions of Type IIB supergravity. 
We construct a null-gauged WZW model whose target-space geometry reproduces this  background in the NS5 decoupling limit, and show that the gauging parameters  precisely  satisfy the appropriate consistency conditions on the perturbative string spectrum. Through an analytic continuation and an appropriate limit, this model relates to a recently studied construction describing strings on smooth, horizonless geometries.
We further analyze the supersymmetric and extremal three-charge configuration, corresponding to the ten-dimensional uplift of the (NS5-decoupled) BMPV black hole,
and show that it requires a novel class of null-gauged WZW models.
Our construction provides explicit examples of worldsheet CFTs that capture string dynamics on certain black hole geometries with linear dilaton asymptotics.
}

\newpage
\setcounter{page}{1}
\pagestyle{plain}

\tableofcontents

\newpage


\section{Introduction}

The D1-D5-P system of Type IIB string theory, and its S-dual description in terms of bound states of NS5-branes, fundamental strings (F1), and momentum modes (P) along a compact direction, has provided a rich laboratory for exploring the quantum properties of black holes. It was in this setting that the first successful counting of black hole microstates was achieved~\cite{Strominger:1996sh}. By taking an appropriate decoupling limit, one of the earliest and most celebrated examples of holographic duality was obtained~\cite{Maldacena:1997re}. 

In this work we consider this brane system focusing on the worldsheet perspective. Our purpose is to construct worldsheet CFTs that provide a description of the doubly-rotating NS5-F1-P black hole, in the NS5-brane decoupling limit. After this decoupling limit the asymptotically flat region of the original solution is excised, and the resulting geometry becomes an asymptotically linear dilaton spacetime~\cite{Aharony:1998ub}.

\medskip

We build on recent progress in understanding a class of three-charge configurations in a regime where the geometry is smooth and horizonless. In particular, work initiated in~\cite{Martinec:2017ztd,Martinec:2018nco} has shown that string theory on certain spectrally flowed circular supertubes admits, in the NS5-brane decoupling limit, an exact worldsheet description in terms of null-gauged Wess-Zumino-Witten models.\footnote{By exact we refer to the underlying gauged WZW model at finite level. Its semiclassical target-space fields are obtained by integrating out the gauge fields and can be directly compared with the corresponding supergravity backgrounds.}
  These solutions, first constructed in supergravity in~\cite{Jejjala:2005yu} and commonly referred to as JMaRT, describe RG flows from the NS5-brane theory in the UV to orbifolds of $\mathrm{AdS}_3$ in the IR.

To construct the null-gauged models, one starts from a WZW theory on the twelve-dimensional group manifold $\mathcal G^{\rm up}= SL(2,\mathbb R) \times SU(2) \times \mathbb R \times U(1) \times T^4 $ which we will refer to as the {\it upstairs} model. In practice, one works with the universal cover of $SL(2,\mathbb{R})$ and considers $\mathrm{AdS}_3$ in global coordinates.
The construction proceeds by gauging two null currents, built as chiral and anti-chiral linear combinations of the Cartan generators of $SL(2,\mathbb{R}) \times SU(2)$ and of the momenta along $\mathbb{R} \times U(1)$. Because the gauging involves null currents, the model is automatically anomaly-free.
The resulting {\it downstairs} theory, obtained after gauging, is thus a critical ten-dimensional string theory whose semiclassical target space reproduces the JMaRT background, including its supersymmetric limit~\cite{Giusto:2004id,Giusto:2004ip,Giusto:2012yz}. Subsequent works explored the perturbative spectrum of propagating strings and D-branes in quite some detail~\cite{Martinec:2018nco,Martinec:2019wzw,Martinec:2020gkv,Bufalini:2021ndn,Dei:2024uyx,Martinec:2025xoy}.

The horizonless background of~\cite{Jejjala:2005yu} plays an important role in the fuzzball program~\cite{Mathur:2005zp,Skenderis:2008qn,Bena:2022rna}\footnote{In particular, circular supertubes have proven to be important seed solutions in the construction of more general families of microstate geometries, see e.g.\cite{Bena:2017xbt}.}. This solution was originally obtained as a special limit of the general asymptotically flat, doubly-rotating, three-charge black hole constructed in~\cite{Cvetic:1996xz} and uplifted to ten dimensions following~\cite{Cvetic:1997uw}. The full geometry depends on six independent parameters, and fixing one of them in a particular way yields the horizonless soliton of~\cite{Jejjala:2005yu} (up to certain quantization conditions on some of the remaining parameters). For generic values of the six parameters, however, the ten-dimensional solution describes a black object with an event horizon -- more precisely, a black string (or black brane) from the point of view of six (or ten) dimensions, although by a slight abuse of terminology we will often refer to it simply as a {\it black hole}. This black hole background thus contains one additional parameter. In this case, the NS5-decoupled solution interpolates between a linear dilaton region in the UV and a BTZ black hole~\cite{Banados:1992wn,Banados:1992gq} in the IR.

\medskip 
In this work we construct a more general coset model for the full parameter space of the three-charge configuration, including the most general black hole solution.
We do so by a suitable modification of the model, by adopting a hyperbolic parametrization of the $SL(2,\mathbb R)$ factor, adapted to the description of non-extremal BTZ backgrounds~\cite{Natsuume:1996ij,Hemming:2001we}, as opposed to the elliptic basis used for AdS$_3$.\footnote{The string spectrum on non-extremal BTZ backgrounds has been extensively studied, for instance in~\cite{Martinec:2001cf,Martinec:2002xq,Troost:2002wk,Hemming:2002kd,Rangamani:2007fz,Ashok:2021ffx,Martinec:2023plo}.} 
As a consequence, the null-gauging is also modified: the procedure now involves the non-compact hyperbolic generator of $SL(2,\mathbb R)$, instead of the compact one as in previous models~\cite{Martinec:2017ztd,Martinec:2018nco}. This has important consequences for the spectrum of propagating strings, some of which we discuss below.

Physical states in the null-gauged WZW model are selected by the holomorphic and antiholomorphic Virasoro constraints, together with a pair of gauge constraints that restrict the allowed quantum numbers~\cite{Martinec:2017ztd}. Analyzing these conditions for representations with arbitrary spectral flow reveals non-trivial restrictions on the a priori continuous gauging parameters that specify how the gauge group $U(1) \times U(1)$ is embedded into ${\cal G}^{\rm up}$. In general, the construction involves eight real gauging parameters which, together with the level of the ${\rm sl}(2,\mathbb R)$ algebra (related to the number of five-branes), form a total of nine parameters characterizing the worldsheet theory. These are subject to three relations: two null constraints, ensuring that the gauge currents are null, as well as an additional constraint ensuring consistent gauge orbits along non-compact directions, leaving six independent parameters, in agreement with the gravitational analysis.

A study of these relations further shows that five of these parameters are quantized, with the associated integers corresponding to the discrete electric charges and angular momenta carried by the brane configuration. Two remaining gauging parameters are mapped to the mass and angular momentum of the non-extremal rotating BTZ black hole in the upstairs model.\footnote{However, the null constraints impose a relation between the BTZ angular momentum and the ten-dimensional black hole conserved charges, reducing the number of independent parameters to six, as expected.}

Moreover, we show that the coset CFT of~\cite{Martinec:2017ztd,Martinec:2018nco,Bufalini:2021ndn}, corresponding to horizonless supertube backgrounds, is recovered by taking the latter two gauging parameters to be purely imaginary and equal. This limit corresponds to sending the inverse Hawking temperature ($\beta$) of the black hole to zero, $\beta \to 0$,\footnote{This observation was inspired by a similar mechanism identified in~\cite{Cassani:2024kjn,Cassani:2025iix} in the context of asymptotically flat supersymmetric solutions of five-dimensional supergravity.} thereby connecting the black hole regime to the horizonless soliton.

\medskip

Another regime of interest is the supersymmetric limit of the three-charge solution of~\cite{Cvetic:1996xz}. In this work we restrict our attention to Lorentzian geometries and find that only two well-behaved supersymmetric backgrounds are possible: the supersymmetric spectrally flowed circular supertube~\cite{Jejjala:2005yu,Giusto:2004id,Giusto:2004ip,Giusto:2012yz} and the BMPV black hole~\cite{Breckenridge:1996is}. The coset CFT corresponding to the former is included in the class studied in~\cite{Martinec:2017ztd} (by imposing a suitable relation among the parameters), while here we focus on the extremal black hole regime.  

In this case, constructing the null-gauged WZW model requires modifying the upstairs $SL(2,\mathbb{R})$ parametrization to one adapted to an extremal rotating BTZ black hole. This is achieved by adopting an \emph{asymmetric} parametrization~\cite{Parsons:2009si}: the holomorphic sector behaves like that of the non-extremal BTZ, requiring a basis that diagonalizes the hyperbolic current ${\cal J}_{\rm sl}^3$, whereas the antiholomorphic sector behaves like the zero-mass extremal BTZ, requiring diagonalization of the parabolic current $\bar{\cal J}_{\rm sl}^-$. Consequently, we are led to a new class of null-gauged models in which the gauged holomorphic current contains ${\cal J}_{\rm sl}^3$, while the gauged antiholomorphic current contains $\bar{\cal J}_{\rm sl}^-$.
%

\medskip

This paper is organized as follows. In section~\ref{sec:sugra} we review the general doubly rotating, three-charge asymptotically flat black hole of~\cite{Cvetic:1996xz,Cvetic:1997uw} and we discuss the NS5 decoupling limit leading to an asymptotically linear dilaton background. We then show that an additional decoupling limit gives the BTZ$\times S^3\times T^4$ near-horizon geometry. 
In section~\ref{sec:worldsheet} we construct the corresponding null-gauged WZW model and derive its target-space geometry. We provide a map between the gauging parameters characterizing the string background and the parameters describing the black hole. We also discuss some properties of the background, such as the absence of closed timelike curves (CTCs), its  Hawking temperature, and the relation of our model to earlier constructions of spectrally flowed supertubes~\cite{Martinec:2017ztd,Martinec:2018nco,Bufalini:2021ndn}. 
Section~\ref{sec:stringspectrum} investigates various features of the string spectrum, showing that consistency of the gauged model imposes four quantization conditions on the gauging parameters and fixes two of them in terms of  the mass and the angular momentum of the BTZ black hole.
Finally, in section~\ref{sec:susy} we consider the background obtained by imposing supersymmetry and extremality. We show that in this limit the worldsheet CFT is determined by a null-gauged WZW model that is adapted to the mixed hyperbolic/parabolic parametrization of the extremal BTZ black hole.
We present our conclusions in sec.~\ref{sec:discussion}.
Three appendices complement our analysis: app.~\ref{app:WZWmodels} reviews gauged and null-gauged WZW models; app.~\ref{app:BTZspectrum} summarizes some features of string theory on BTZ and its spectral flow; app.~\ref{sec:CTC} shows that, after imposing supersymmetry, there exist only two consistent Lorentzian backgrounds: the supersymmetric spectrally flowed circular supertube~\cite{Giusto:2004id,Giusto:2004ip,Giusto:2012yz} and the BMPV black hole~\cite{Breckenridge:1996is}. These correspond to opposite limits of the inverse Hawking temperature.


\section{Supergravity solutions}
\label{sec:sugra}

In this section we consider the supergravity solution describing the asymptotically flat, doubly-rotating, three-charge black hole first derived in~\cite{Cvetic:1996xz}, and uplifted to ten-dimensional supergravity following~\cite{Cvetic:1997uw}. This black hole provides the low-energy effective description of a bound state of D-branes in Type IIB superstring theory compactified on ${S}^1 \times T^4$ involving $n_5$ D5-branes, wrapping both the ${S}^1$ and the four-torus $ T^4$, as well as $n_1$ D1-branes wrapping ${ S}^1$ and smeared along $ T^4$, carrying an additional momentum charge $n_P$ along the compact ${S}^1$. In the S-dual frame the underlying system consists of bound states of NS5-branes and F1-strings with momentum charge.  Our focus will be on the properties of the geometry emerging in the near-horizon region of the five-branes. Indeed, a suitable NS5-decoupling limit removes the asymptotically flat region and the corresponding solution has an asymptotically linear dilaton behaviour.

A similar analysis was performed in~\cite{Martinec:2018nco} starting from the asymptotically flat horizonless solution of~\cite{Jejjala:2005yu}. This solution, usually referred to as the JMaRT soliton, is a specific limit of the general NS5-F1-P system, where a certain combination of the two angular momenta is fixed in terms of the other conserved charges. Here, we will be interested in the regime of parameters in which the supergravity solution describes a black hole with a Killing horizon and finite entropy, and we will highlight key differences with respect to the JMaRT case. 

\medskip

The string frame metric for the brane system mentioned above in the coordinates $(r,t,\theta,\phi,\psi,y,z_a)$ is given by
\begin{equation}
\label{eq:ns5f1pmetric}
\begin{aligned}
{\rm d}s^2 &= -\frac{f}{\tilde H_1}\left( {\rm d} t^2 - {\rm d} y^2\right) + \frac{M}{\tilde H_1}\left( c_p\,{\rm d}t - s_p\,{\rm d}y\right)^2 + \tilde H_5 \left( \frac{r^2 \,{\rm d}r^2}{\Delta_r} + {\rm d}\theta^2\right) + \frac{M}{\tilde H_1}\nu^2\\
&+\left( \tilde H_5 - \left( a_2^2 - a_1^2\right) \frac{\tilde H_1 + \tilde H_5 - f}{\tilde H_1}\cos^2\theta\right) \cos^2\theta\,{\rm d}\psi^2 + \frac{2M}{\tilde H_1} \left( \tilde\gamma_1\, {\rm d}t + \tilde \gamma_3\, {\rm d}y\right)\cos^2\theta\,{\rm d}\psi
\\
&+ \left( \tilde H_5 + \left( a_2^2 - a_1^2\right) \frac{\tilde H_1 + \tilde H_5 - f}{\tilde H_1}\sin^2\theta\right) \sin^2\theta\,{\rm d}\phi^2 + \frac{2M}{\tilde H_1} \left( \tilde\gamma_2\, {\rm d}t + \tilde \gamma_4\, {\rm d}y\right)\sin^2\theta\,{\rm d}\phi\\
&+ \sum_{a=1}^4{\rm d}z_a^2\,,
\end{aligned}
\end{equation}
where
\begin{equation}
\begin{aligned}
\nu &= a_1 \cos^2\theta\,{\rm d}\psi + a_2 \sin^2\theta\,{\rm d}\phi\,,\\[1mm]
f &= r^2 + a_1^2 \sin^2\theta + a_2^2 \cos^2\theta\,,\qquad \Delta_r = \left( r^2 + a_1^2 \right)\left( r^2 + a_2^2\right) - M r^2\,,\\[1mm]
\tilde H_1 &= f + M s_1^2\,,\qquad \tilde H_5 = f + M s_5^2\,,\qquad s_i \equiv \sinh\delta_i\,,\qquad c_i \equiv \cosh\delta_i\,,\,\,\text{with} \,\, i = 5,\,1,\,p,\\[1mm]
\tilde \gamma_1 &= a_1 c_1 c_5 c_p - a_2 s_1 s_5 s_p\,,\qquad \tilde \gamma_2 = a_2 c_1 c_5 c_p - a_1 s_1 s_5 s_p\,,\\[1mm]
\tilde\gamma_3 &= a_2 s_1 s_5 c_p - a_1 c_1 c_5 s_p\,,\qquad \tilde \gamma_4 = a_1 s_1 s_5 c_p - a_2 c_1 c_5 s_p\,.
\end{aligned}
\end{equation}
Here, $y\sim y + 2\pi R_y$ parametrizes the compact direction ${S}^1$, while $z_a$, with $a=1,2,3,4$, are coordinates on $ T^4$.
The $T^4$ factor will play almost no role in the following discussion, and we will neglect it most of the time to keep the notation simpler. As a consequence, the metric \eqref{eq:ns5f1pmetric} is effectively regarded as a six-dimensional solution. 

The background also entails a non-trivial dilaton $\Phi$ and an NS-NS B-field,
\begin{equation}
\begin{aligned}
B_2 &= \frac{M}{\tilde H_1}\cos^2\theta\Bigl[ \left( a_2 c_1 s_5 c_p - a_1 s_1 c_5 s_p\right) {\rm d}t + \left( a_1 s_1 c_5 c_p - a_2 c_1 s_5 s_p\right) {\rm d}y\Bigr]\wedge {\rm d} \psi\\
&+\frac{M}{\tilde H_1}\sin^2\theta\Bigl[ \left( a_1 c_1 s_5 c_p - a_2 s_1 c_5 s_p\right) {\rm d}t + \left( a_2 s_1 c_5 c_p - a_1 c_1 s_5 s_p\right) {\rm d}y\Bigr]\wedge {\rm d} \phi\\
&- \frac{M}{\tilde H_1}s_1 c_1 \,{\rm d}t \wedge {\rm d}y + \frac{M}{\tilde H_1}s_5 c_5\left( r^2 + a_2^2 + M s_1^2\right) \cos^2\theta\,{\rm d}\phi \wedge {\rm d}\psi\,,
\\[1mm]
{\rm e}^{2\Phi} &=g_{\rm s}^2\frac{\tilde H_5}{\tilde H_1} \,.
\end{aligned}
\end{equation}
The black hole depends on six independent parameters, $(M, \delta_{1,5,p}\,, a_{1,2})$, which control the six conserved charges, denoted respectively as $(E, Q_1, Q_5, Q_p, J_1, J_2)$. The map between conserved charges and independent parameters is given by
\begin{equation}
\begin{aligned}
E &= \frac{\pi}{8G_{(5)}}M\left( c_1^2 + c_5^2 + c_p^2 + s_1^2 + s_5^2 + s_p^2\right)\,,\\[1mm]
Q_i &= \frac{\pi}{4G_{(5)}}M s_i c_i\,,\quad J_1 = -\frac{\pi}{4G_{(5)}}M \tilde \gamma_1 \,,\quad J_2= -\frac{\pi}{4G_{(5)}} M \tilde \gamma_2\,,
\end{aligned}
\end{equation}
where $G_{(5)}$ is the five-dimensional Newton's constant
\begin{equation}
G_{(5)} = \frac{G_{(10)}}{2\pi R_y \,V_{T^4}}\,, \qquad\qquad G_{(10)} = 8\pi^6g_{\rm s}^2 \,\ell_{\rm s}^8\,,
\end{equation}
being $g_{\rm s}$ and $\ell_{\rm s}= \sqrt{\alpha'}$ the string coupling and string length, respectively. In the following, we will take the volume of the $T^4$ to be $V_{T^4}= \left( 2\pi\right)^4 v\,\ell_{\rm s}^4$. Without loss of generality we can take $M\geq 0$, $\delta_{1,5,p}\geq 0$ and $a_1 \geq a_2 \geq 0$. These parameters are related to the integers $n_{1,5,P}$ by
\begin{equation}
\label{eq:charges_params_string}
Ms_1c_1 = \frac{g_{\rm s}^2\,\ell_{\rm s}^2}{v}n_1\,,\qquad M s_5 c_5 = \ell_{\rm s}^2n_5\,,\qquad M s_p c_p = \frac{g_{\rm s}^2\,\ell_{\rm s}^4}{R_y^2\,v}n_P\,.
\end{equation}

The event horizon is located at the largest positive root of the equation $\Delta_r = 0$, denoted by $r_+$. This equation is solved by the roots
\begin{equation}
r_\pm^2= \frac{1}{2}\left(M -a_1^2 - a_2^2 \pm \sqrt{\left( a_1^2 + a_2^2 -M\right)^2 -4 a_1^2 a_2^2 }\right)\,,
\end{equation}
which are real as long as $|M- a_1^2 - a_2^2| > 2 a_1a_2$. Depending on the possible values of the parameter 
$M$, this inequality can be satisfied in two different ways, corresponding to two distinct branches of the solution~\cite{Jejjala:2005yu}:

\begin{enumerate}
\item If 
\begin{equation}
M \geq \left( a_1 + a_2\right)^2\,,
\end{equation}
the hypersurface at $r=r_+$ is an event horizon, and the solution describes a black hole. This is the regime of primary interest in this work. The black hole reaches the extremal limit as $r_-\rightarrow r_+$, which corresponds to saturating the above inequality.
\item Horizonless solutions can be obtained in the other case: 
\begin{equation}
\label{eq:rangeMsoliton}
M < \left( a_1 - a_2\right)^2\,.
\end{equation}
To obtain a causally well-behaved horizonless soliton (smooth up to possible discrete orbifold singularities) one should also impose that a certain spacelike compact direction contracts to zero at $r=r_+$, where the geometry closes. This condition translates into
\begin{equation}
\label{eq:JMaRT}
M = a_1^2 + a_2^2 - a_1 a_2 \frac{c_1^2 c_5^2 c_p^2 + s_1^2 s_5^2 s_p^2}{s_1 c_1 s_5 c_5 s_p c_p}\,.
\end{equation}
We can interpret this as imposing a 
relation between the conserved charges of the solution, so that the horizonless soliton only has five independent charges. The global structure of the resulting geometry fixes the radius of the $S^1$ parametrized by $y$ in terms of the other parameters of the solution. It also introduces three quantization conditions, specifying the orbifold structure of the geometry.
Since our primary focus is on the black hole regime, we will not elaborate on these conditions here, and refer the reader to the original reference for more details. 
\end{enumerate}

Before discussing the five-brane decoupling limit, it is useful to introduce a more natural parametrization, which simplifies the process of taking various limits. To this end, we define a new set of six independent real parameters, $(M, q_{1,5,p}\,,\ell \,,b)$ given by
\begin{equation}
q_i = \frac{M}{4}e^{2\delta_i}\,, \qquad b= \sqrt{M}(a_1 + a_2) \,,\qquad \ell = \frac{a_1 - a_2}{\sqrt{M}}  \,.
\end{equation}
The locations of the outer and inner horizons are now given by
\begin{equation}
\label{eq:horizonlocation}
r_\pm^2 =\frac{M}{2}- \frac{b^2}{4M}- \frac{M}{4}\ell^2\pm \frac{\sqrt{\left( 1- \ell^2\right)\left( M^2 - b^2\right)}}{2}\,,
\end{equation}
In terms of the new parameters, the branch of the solution corresponding to the black hole is characterized by
\begin{equation}
M^2 - b^2 \geq 0\,.
\end{equation}
The outer horizon \eqref{eq:horizonlocation}, then, remains real as long as $0\leq \ell<1$. The extremality condition is translated into the limit $b\rightarrow M$. Finally, in this parametrization, the conserved charges carried by the black hole are expressed as
\begin{equation}
\begin{aligned}
 E &=\frac{\pi}{4G_{(5)}}\left[ q_1 + q_5 + q_p + \frac{M^2}{16}\left(\frac{1}{q_1}+ \frac{1}{q_5}+ \frac{1}{q_p}\right)\right]\,,\qquad Q_i = \frac{\pi}{4G_{(5)}}\left(q_i - \frac{M^2}{16 q_i}\right) \,,\\[1mm]
\qquad J_+ &\equiv -\left(J_1 + J_2\right) = \frac{\pi}{128G_{(5)}}\frac{b}{\sqrt{q_1 q_5 q_p}}\Bigl(M^2 + 16\left( q_1 q_5 + q_1 q_p + q_5 q_p\right)\Bigr)\,,\\[1mm]
J_- &\equiv  J_2 - J_1 = \frac{\pi}{32G_{(5)}}\frac{\ell}{\sqrt{q_1 q_5 q_p}}\Bigl(M^2 \left( q_1 + q_5 + q_p\right) + 16 q_1 q_5 q_p \Bigr)\,.
\end{aligned}
\end{equation}

On the other hand, the regime in which the solution describes a soliton, as given by \eqref{eq:rangeMsoliton}, corresponds in our choice of parameters to the complementary range of values for $\ell$, namely
\begin{equation}
\ell \geq 1\,,
\end{equation}
with $b>M$ as required by reality of \eqref{eq:horizonlocation}.


\subsection{Five-brane decoupling limit}
\label{sec:decoupling}

A decoupled theory of five-branes can be obtained by taking the limit~\cite{Aharony:1998ub}
\begin{equation}\label{ns5_dec}
g_{\rm s} \to 0\,,\qquad \qquad \frac{M}{g_{\rm s}^2\ell_{\rm s}^2}  = \text{fixed}\,,
\end{equation}
and focusing on radial distances of the order $g_{\rm s} \ell_{\rm s}$. As seen from \eqref{eq:charges_params_string} this is equivalent to taking ${\rm e}^{\delta_5} \to \infty$, while keeping ${\rm e}^{\delta_{1,p}}$ and $q_5$ fixed. In particular, in this limit we find $q_5 = \ell_{\rm s}^2 \,n_5$.  Concretely, the decoupling limit can be taken by replacing~\cite{Martinec:2018nco} 
\begin{equation}\label{ns5_dec_1}
r \to \epsilon\, r\,,\,\qquad\, g_{\rm s} \to \epsilon\,g_{\rm s}\,,\,\qquad\,  M \to \epsilon^2 M\,,\,\,\qquad \, b \to \epsilon^2 b\,,\,\qquad \, q_{1,p} \to \epsilon^2 q_{1,p}\,,
\end{equation}
and then taking the $\epsilon \to 0$ limit. From now on, unless specified differently, we adopt conventions on the parameters such that $\alpha' = \ell_{\rm s}^2 = 1$, $g_{\rm s}=1$ and $G_{(5)} = \pi/4$. 

The decoupled geometry can be expressed in terms of a new radial coordinate defined by
\begin{equation}
\label{eq:def_rho}
\sinh^2\rho = \frac{r^2-r_+^2}{r_+^2 - r_-^2}\,,
\end{equation}
adapted to the worldsheet analysis that will follow later. Note that this radial coordinate is ill-defined in the limit $r_+ \rightarrow r_-$. For this reason in this section we focus on solutions with $\ell \neq 1$ and $b \neq M$. 
After the limit, the metric can be expressed as
\begin{equation}
\label{eq:decoupledmetric}
\begin{aligned}
{\rm d}s^2 &= \Sigma_0^{-1}\left(-\tilde h_{tt}{\rm d}t^2 + \tilde h_{yy}{\rm d}y^2\right) + 2q_1\left( M^2 - 16 q_p^2\right)\Sigma_0^{-1}\,{\rm d}t \,{\rm d}y \\[1mm]
&\quad + q_5 \left(\diff \rho^2+ {\rm d}\theta^2\right)+q_5\,\Sigma_0^{-1}\Bigl[ \tilde h_{\phi\phi} \,\sin^2\theta \,{\rm d}\phi^2 +\tilde h_{\psi\psi} \,\cos^2\theta \,{\rm d}\psi^2\Bigr] \\[1mm]
&\quad + \Sigma_0^{-1}\Bigl[\sin^2\theta\,\left(\tilde \eta_- \, {\rm d}t + \tilde \zeta_-\,{\rm d}y\right){\rm d}\phi +\cos^2\theta\,\left(\tilde \eta_+ \, {\rm d}t + \tilde \zeta_+\,{\rm d}y\right){\rm d}\psi
\Bigr]\,,
\end{aligned}
\end{equation}
where 
\begin{equation}
\label{eq:decoupledmetric1}
\begin{aligned}
\Sigma_0&= 16 \,q_1 \,q_p \,\tilde H_1=\\
&= q_p \left[ M^2 + 16 \,q_1^2 + 8\,q_1\sqrt{\left(1-\ell^2 \right)\left(M^2-b^2 \right)}\cosh2\rho - 8 \,b\,\ell\,q_1\cos2\theta\right]\,,\\[1mm]
-\tilde h_{tt}&= q_1 \left[M^2 + 16 \,q_p^2 -8\,q_p\sqrt{\left(1-\ell^2 \right)\left(M^2-b^2 \right)}\cosh2\rho+ 8 \,b\,\ell \,q_p\cos2\theta\right]
\,,\\[1mm]
\tilde h_{yy}&= q_1 \left[M^2 + 16 \,q_p^2 +8\,q_p\sqrt{\left(1-\ell^2 \right)\left(M^2-b^2 \right)}\cosh2\rho- 8 \,b\,\ell \,q_p\cos2\theta\right]
\,,\\[1mm]
\tilde h_{\psi\psi}&= q_p \left[M^2 + 16 \,q_1^2 + 8\,q_1\sqrt{\left(1-\ell^2 \right)\left(M^2-b^2 \right)}\cosh2\rho+ 8 \,b\,\ell\,q_1\right]
\,,\\[1mm]
\tilde h_{\phi\phi}&=q_p \left[M^2 + 16 \,q_1^2 + 8\,q_1\sqrt{\left(1-\ell^2 \right)\left(M^2-b^2 \right)}\cosh2\rho- 8 \,b\,\ell\,q_1\right]
\,,\\[1mm]
\tilde \eta _\pm &= 2\sqrt{q_1 \,q_5\,q_p} \left[4b(q_1 + q_p)\pm \ell(M^2 + 16 q_1 q_p)\right]\,,\\[1mm]
\tilde \zeta _\pm &= 2\sqrt{q_1 \,q_5\,q_p}\left[4b(q_1 - q_p) \pm \ell (M^2 - 16 q_1 q_p)\right]\,.
\end{aligned}
\end{equation}
The dilaton becomes
\begin{equation}
{\rm e}^{2\Phi} = \frac{16\,q_1\,q_5\,q_p}{\Sigma_0}\,,
\end{equation}
while the B-field is given by
\begin{equation}
\begin{aligned}
B&= \frac{q_p\left( M^2 - 16 q_1^2\right)}{ \Sigma_0}\,\diff t\wedge \diff y + \frac{q_5\,\tilde h_{\phi\phi}}{\Sigma_0}\cos^2\theta\, \diff \phi \wedge \diff \psi\\
&\quad +\frac{1}{2\Sigma_0}\left[ \tilde \eta_+\,\diff t + \tilde\zeta_+\,\diff y\right]\wedge \sin^2\theta\diff\phi + \frac{1}{2\Sigma_0}\left[\tilde \eta_- \,\diff t+ \tilde \zeta_- \,\diff y\right] \wedge \cos^2\theta \diff\psi\,.
\end{aligned}
\end{equation}
Finally, the charges of the solution are now given by\footnote{Note that, in the decoupling limit, the right-hand side in the expression for the mass contributes only at subleading order.}
\begin{equation}
\label{eq:bhcharges}
\begin{aligned}
E &-Q_5 = Q_1 + Q_p + \frac{M^2\left( q_1 + q_p\right)}{8 q_1\, q_p}\,,\\[2mm]
Q_5 &= q_5,\, \qquad Q_{1,p}= q_{1,p}- \frac{M^2}{16 q_{1,p}}\,,\\[2mm]
J_+ &= \frac{b}{2}\sqrt{\frac{q_5}{q_1\,q_p}}\left( q_1 + q_p\right)\,,\qquad J_- = \frac{\ell}{8}\sqrt{\frac{q_5}{q_1\,q_p}}\left( M^2 + 16 q_1 \,q_p\right)\,.
\end{aligned}
\end{equation}
Since the asymptotically flat region has been decoupled from the geometry, the solution has now the asymptotic behaviour typical of the near-horizon region of NS5-branes. Indeed, for large $\rho$, the metric asymptotically tends to
\begin{equation}
\label{eq:lineardilatonbkgr}
\begin{aligned}
{\rm d}s^2 &\,\rightarrow\, - {\rm d}t^2 + {\rm d}y^2 + q_5 \Bigl({\rm d}\rho^2 + \left({\rm d}\theta^2  + \sin^2\theta\, {\rm d}\phi ^2 + \cos^2\theta \,{\rm d}\psi^2\right)\Bigr) \,,
\\[1mm]
\Phi & \,\rightarrow \, -\rho\,.
\end{aligned}
\end{equation}
The decoupled geometry, therefore, gives an asymptotically linear dilaton background. 

\medskip

In this work, we are primarily interested in studying the regime where the solution describes a black hole with horizon, which holds as long as $0\leq \ell <1$. However, for future reference, we find it useful to provide the NS5-decoupled version of the relation \eqref{eq:JMaRT}, which characterizes solitonic solutions. According to our conventions, obtaining a horizonless configuration in the decoupling limit requires imposing
\begin{equation}
\label{eq:decsoliton}
b^2 = M^2 + \left( \ell^2 -1\right)\left(\frac{M^2 + 16 q_1 q_p}{4\left( q_1 + q_p\right)}\right)^2\,,
\end{equation}
into \eqref{eq:decoupledmetric}. Since $\ell>1$ in this regime, a real solution for $b$ always exists.


\subsection{AdS decoupling limit}
\label{sec:F1decoupling}

To reach the AdS$_3\times S^3$ near-horizon region of the NS5-F1-P system a further limit is required, corresponding to the decoupling of the F$1$-string. Then, the asymptotic linear dilaton background is removed from the geometry, giving the asymptotically AdS$_3$ $\times$ S$^3$ throat. In the parametrization we are employing, this limit is obtained by
\begin{equation}
R_y \to +\infty\,,\qquad \qquad MR_y^2 = \text{fixed}\,,
\end{equation}
while also keeping $q_5$ fixed. As seen from \eqref{eq:charges_params_string}, this is equivalent to taking ${\rm e}^{\delta_1} \to +\infty$, while keeping ${\rm e}^{\delta_p}$ and $q_1$ fixed. 

One can take this limit in the solution above by first introducing the new coordinates
\begin{equation}
\label{eq:change_coord_ads}
t = R_y\, \hat t\,, \qquad \qquad  y = R_y \, \hat y\,, \qquad \qquad \hat R^2 = \left( r_+^2 - r_-^2\right) \sinh^2\rho\,,
\end{equation}
making the substitutions
\begin{equation}
\hat R \to \epsilon\,\hat R\,, \qquad R_y \to \frac{R_y}{\epsilon}\,,\qquad M \to \epsilon^2M\,,\qquad b \to \epsilon^2 b\,,\qquad q_p \to \epsilon^2 q_p\,,
\end{equation}
and then taking the limit $\epsilon \to 0$. After the decoupling limit the six-dimensional metric takes the form of a product of a three-sphere and a three-dimensional black hole with AdS$_3$ asymptotics, i.e.~a rotating BTZ black hole~\cite{Banados:1992wn,Banados:1992gq}, with the three-sphere fibered over the BTZ factor~\cite{Cvetic:1998xh}:
\begin{equation}
\begin{aligned}
\diff s^2&=q_5\Bigl\{ -\frac{\left( \hat r^2 - \hat r_+^2\right)\left( \hat r^2 - \hat r_-^2\right)}{\hat r^2} \diff \hat t^2+ \frac{\hat r^2}{\left( \hat r^2 - \hat r_+^2\right)\left( \hat r^2 - \hat r_-^2\right)}\diff \hat r^2 + \hat r^2\left( \diff \hat y - \frac{\hat r_+\,\hat r_-}{\hat r^2}\diff \hat t\right)^2
\\[2mm]
&+ \diff \theta^2 +\sin^2\theta\left( {\rm d}\phi + W_-\, {\rm d}\hat t + W_+ \, {\rm d} \hat y\right)^2 +\cos^2\theta\left( {\rm d}\psi + W_+\, {\rm d} \hat t + W_- \, {\rm d}\hat y\right)^2\Bigr\}\,,
\end{aligned}
\end{equation}
where
\begin{equation}
\label{eq:functionsAdS3}
\begin{aligned}
\hat R^2 &= q_1 q_5\left(\hat r^2 - \hat r_+^2\right)\,,\qquad \hat r_\pm^2 = \frac{M^2 - b^2 +16\left( 1- \ell^2\right) q_p^2}{16 q_1 q_5 q_p} \pm \frac{1}{2q_1 q_5} \sqrt{\left( 1-\ell^2\right)\left(M^2-b^2\right)}\,,
\\[1mm]
W_\pm &= \frac{J_+ \pm J_-}{2\,q_1\,q_5}\,,
\end{aligned}
\end{equation}
and the dilaton becomes constant
\begin{equation}
{\rm e}^{2\Phi}= \frac{q_5}{q_1}\,.
\end{equation}
The charges appearing in \eqref{eq:functionsAdS3} are those obtained after the limit, given by
\begin{equation}
\begin{aligned}
E&-\left( Q_1 + Q_5 + Q_p\right) = \frac{M^2}{8\,q_p}\,,\qquad Q_{1,5} = q_{1,5}\,,\qquad Q_p = q_p - \frac{M^2}{16 q_p}\,,\\
J_- &= 2\ell \sqrt{Q_1\,Q_5\,q_p}\,,\qquad J_+ = \frac{b}{2}\sqrt{\frac{Q_1\,Q_5}{q_p}}\,.
\end{aligned}
\end{equation}

\medskip

Before concluding this section, let us comment about the F$1$-decoupling limit in the soliton regime, where $\ell>1$. To do so, we first reintroduce the coordinate $\rho$ of \eqref{eq:change_coord_ads}, and then we impose the constraint \eqref{eq:JMaRT}, which now takes the form
 \begin{equation}
 b = \sqrt{M^2 + 16\left( \ell^2-1\right)q_p^2}\,.
\end{equation}
Then, one finds that the metric becomes~\cite{Martinec:2018nco}
\begin{equation}
\begin{aligned}
\diff s^2 &= \frac{4\left(\ell^2-1\right)q_p}{q_1}\Bigl[ - \cosh^2\rho \,\diff\hat t^2 + q_5\,\diff \rho^2 + \sinh^2\rho \,\diff\hat y^2 \Bigr] 
\\[1mm]
&+q_5\Bigl[ \diff \theta^2 +\sin^2\theta\left( {\rm d}\phi + W_-\, {\rm d}\hat t + W_+ \, {\rm d} \hat y\right)^2 +\cos^2\theta\left( {\rm d}\psi + W_+\, {\rm d} \hat t + W_- \, {\rm d}\hat y\right)^2\Bigr]\,.
\end{aligned}
\end{equation}

This highlights a key distinction between the black hole and horizonless regimes: in the latter case we obtain a global $AdS_3$ factor as the near-horizon limit of the JMaRT soliton, as in~\cite{Maldacena:2000dr}. 
Depending on the values of the parameters $W_{\pm}$, this solution in fact describes orbifolds of $AdS_3 \times S^3$ whose detailed analysis can be found, for instance, in~\cite{Martinec:2018nco}.


\subsection{Non-rotating solution}
\label{sec:mald_example}

In this section we focus on the non-extremal, non-rotating three-charge black hole, whose ten-dimensional embedding was analyzed in~\cite{Callan:1996dv,Maldacena:1996ky}, obtained from the broader class of solutions discussed above by turning off the angular momenta.
Our goal is to study the thermodynamics of the non-extremal solution in the NS5 decoupling limit in this simpler setting. Many expressions simplify considerably, enabling us to present explicit and compact formulas for the temperature and entropy of the solution.

In the NS5-F1-P frame the metric \eqref{eq:ns5f1pmetric} with $a_1 = a_2 =0$ can be expressed as 
\begin{equation}
\label{eq:Mald_blackhole}
\begin{aligned}
\diff s^2 = f_1^{-1}\left[ - f_p^{-1}f_0 \diff t^2 + f_p \left( \diff y + \frac{M s_p c_p}{f_p r^2}\diff t\right)^2\right] + f_5 \left( f_0^{-1} \diff r^2 + r^2 \diff \Omega_3^2\right) + \diff s^2_{T^4}\,,
\end{aligned}
\end{equation}
where 
\begin{equation}
f_0 = 1- \frac{M}{r^2}\,,\qquad f_i = 1 + \frac{M s_i^2}{r^2}\,,\quad i = 1,5,p\,.
\end{equation}

The geometry possesses a Killing horizon located at $r_+ = \sqrt{M}$, generated by the Killing vector
\begin{equation}
\xi = \partial_t + \frac{s_p}{c_p}\partial_y\,,
\end{equation}
whose norm $\xi^\mu\xi_\mu$ vanishes for $r\to r_+$. The inverse Hawking temperature of the black hole is given by
\begin{equation}
\beta = 2\pi \sqrt{M}\, c_1\,c_5\,c_p = \frac{\pi}{32M}\frac{\left( M + 4 q_1\right)\left( M + 4 q_5\right)\left( M + 4 q_p\right)}{\sqrt{q_1 q_5 q_p}}\,,
\end{equation}
while Bekenstein-Hawking entropy reads 
\begin{equation}
\label{eq:Mald_entropy}
{\cal S} = \frac{R_y v}{g_{\rm s}^2\ell_{\rm s}^4}M\beta\,.
\end{equation}
Note that in this section we have reintroduced the explicit dependence on the string coupling, the string length and the gravitational constant.

In this simple example the NS5 decoupling limit \eqref{ns5_dec} and \eqref{ns5_dec_1} amounts to dropping the $1$ in the harmonic function $f_5$ of \eqref{eq:Mald_blackhole}, obtaining 
\begin{equation}
\label{eq:Mald_blackhole_dec}
\begin{aligned}
\diff s^2 = f_1^{-1}\left[ - f_p^{-1}f_0 \diff t^2 + f_p \left( \diff y + \frac{M s_p c_p}{f_p r^2}\diff t\right)^2\right] + \frac{\ell_{\rm s}^2}{r^2}n_5 \left( f_0^{-1} \diff r^2 + r^2 \diff \Omega_3^2\right) + \diff s^2_{T^4}\,,
\end{aligned}
\end{equation}
where we used that $q_5 = \ell_{\rm s}^2 n_5$ in this limit. After the decoupling the solution retains a finite inverse temperature
\begin{equation}
\beta = \frac{\pi \ell_{\rm s}}{8M}\sqrt{\frac{n_5}{q_1\,q_p}}\left( M+ 4 q_1\right)\left( M+ 4 q_p\right)\,,
\end{equation}
and an associated entropy still given by \eqref{eq:Mald_entropy}.

After the NS5 decoupling, the entropy \eqref{eq:Mald_entropy} interpolates between the Cardy entropy of a BTZ black hole in the IR and a Hagedorn growth of states at high energies~\cite{Chakraborty:2020swe,Apolo:2019zai}.\footnote{The notation of~\cite{Chakraborty:2020swe} can be mapped to the one employed here by taking $r_0^2 \to M$, $\alpha_i \to \delta_i$, $p\to n_1$, $k \to n_5$, $n\to n_P$.} This analysis matches the predictions for the spectrum and degeneracy of states in a $T\bar T$-deformed CFT~\cite{Smirnov:2016lqw,Cavaglia:2016oda}, which displays the same interpolation (see also~\cite{Giveon:2017nie,Giveon:2017myj,Chakraborty:2020yka} for related discussions). The background \eqref{eq:Mald_blackhole_dec}, then, appears to provide a bulk realization for a family of (thermal) $T\bar T$-deformed CFTs: indeed, in the IR it reduces to a BTZ black hole (via the AdS decoupling limit described above), while in the UV it approaches a linear dilaton space, naturally associated with Hagedorn growth~\cite{Maldacena:1996ya,Kutasov:2000jp}. As we will show in the next section, this solution belongs to the class of backgrounds for which we provide a worldsheet CFT in terms of a null-gauged WZW model.\footnote{Alternative worldsheet models for such a background, but neglecting the three-sphere, have been proposed in~\cite{Giveon:2003ge,Giveon:2005mi,Chakraborty:2020yka}. More general irrelevant deformations that also involve the three-sphere have been studied in~\cite{Apolo:2021wcn,Chakraborty:2022dgm}.}



\section{Black hole CFT}
\label{sec:worldsheet}

In this section we build on the results of~\cite{Martinec:2017ztd,Martinec:2018nco,Martinec:2019wzw,Martinec:2020gkv,Bufalini:2021ndn} and introduce the worldsheet CFT whose target space, in the supergravity regime, reproduces the NS5-F1-P background in the five-brane decoupling limit discussed above. These CFTs arise as null-gauged WZW models
\begin{equation}
\frac{SL(2,\mathbb R) \times SU(2) \times \mathbb R \times U(1)}{U(1)_L \times U(1)_R} \times T^4\,.
\end{equation}
The upstairs group
\begin{equation}
\label{eq:upgroup}
{\cal G}^{\rm up} ={ SL}(2,\mathbb R) \times SU(2) \times \mathbb R \times U(1)\times T^4\,,
\end{equation}
involves the universal cover of $SL(2,\mathbb R)$, an $SU(2)$ factor spanning a three-sphere, $\mathbb R$ denoting an auxiliary timelike direction parametrized by a non-compact coordinate $t$, as well as an auxiliary spatial circle $S^1$ with periodic coordinate $y$. As we will show, there exists a suitable embedding of the subgroup $U(1)_L \times U(1)_R$ into ${\cal G}^{\rm up}$ such that the corresponding coset theory provides a worldsheet description of the black hole background \eqref{eq:decoupledmetric}-\eqref{eq:decoupledmetric1}. More precisely, this construction gauges an $\mathbb R \times U(1)$ subgroup, thereby removing one timelike and one spacelike direction from the twelve-dimensional upstairs group manifold. The $T^4$ factor is unaffected by the gauging and does not play a significant physical role in this context, so we will omit it in the following discussion. For a review of general aspects of gauged WZW models, see appendix~\ref{app:WZWmodels}.

A main difference from previous works lies in the global identifications of the model. In the region of parameter space where the NS5-F1-P background develops a horizon, the IR decoupling limit yields a BTZ factor in the near-horizon of the full brane system, rather than global $AdS_3$. This motivates us to adopt the hyperbolic parametrization of $SL(2,\mathbb R)$~\cite{Natsuume:1996ij},
\begin{equation}
g_{\rm sl} = {\rm e}^{\frac{\sigma + \tau}{2}\sigma_3}e^{\rho \sigma_1} {\rm e}^{\frac{\sigma-\tau}{2}\sigma_3}\,,
\end{equation}
where $\sigma_i$ denote the Pauli matrices.\footnote{Explicitly, they are given by
\begin{equation}
\sigma_1 = \begin{pmatrix} 0 & 1\\1& 0\end{pmatrix}\,,\qquad \sigma_2 = \begin{pmatrix} 0 & -i \\ i & 0\end{pmatrix} \,,\qquad \sigma_3 = \begin{pmatrix} 1 & 0 \\ 0 & -1 \end{pmatrix}\,.
\end{equation}}
In this parametrization the associated bi-invariant metric takes the form
\begin{equation}
\label{eq:BTZ_metric}
\diff s^2_{\rm BTZ} =\mathtt k_{\rm sl}\Bigl( -\sinh^2\rho\, \diff \tau^2 + \diff \rho^2 + \cosh^2\rho\, \diff \sigma^2 \Bigr)\,,
\end{equation}
where $\mathtt k_{\rm sl}$ is the level of the associated Kac-Moody algebra. This describes the BTZ black hole after appropriate identifications.\footnote{See Appendix~\ref{app:BTZspectrum} for more details.} 
Locally, this is related to AdS$_3$ in elliptic coordinates by the analytic continuation
\begin{equation}\label{eq:btztoads}
\tau = -i \sigma_{{\rm AdS}3}, \qquad \sigma = i \tau_{{\rm AdS}_3}\,,
\end{equation}
but the two geometries differ significantly in their global structure. As we will see in the next section, this distinction has important consequences for the null-gauged model.

\subsection{The upstairs model}
\label{sec:upstairs}

We start by choosing a parametrization for the upstairs group elements $g \in \mathcal G^{\rm up}$, with $\mathcal G^{\rm up}$ given by \eqref{eq:upgroup} (neglecting the $T^4$):
\begin{equation}
\label{eq:upstairsgroupelement}
g = \left(e^{\frac{1}{2}(\tau + \sigma)\sigma_3}e^{\rho\,\sigma_1}e^{\frac{1}{2}(\sigma - \tau)\sigma_3}\,,\,e^{\frac{i}{2}(\psi - \phi)\sigma_3}e^{i\theta\,\sigma_1}e^{\frac{i}{2}(\psi + \phi)\sigma_3}\,,\,e^t \,,\,e^{i \frac{y}{R_y}} \right)\,,
\end{equation} 
where $\theta \in [0; \pi/2]$, $\phi$ and $\psi$ are $2\pi$-periodic, and $y \sim y + 2\pi R_y$. 
The WZW model on $\mathcal G^{\rm up}$, parametrized by $g$, is described by the action
\begin{equation}
\label{eq:upWZWmodel0}
S_{WZW} =\sum_I {\rm sgn}\frac{\mathtt k_I}{2\pi}\left( \int_{\Sigma_2}{\rm d}^2z\,\text{Tr}\left[ \partial g\, g^{-1} \overline \partial g\,g^{-1}\right]+ \frac{1}{3}\int_{\Omega_3}\text{Tr}\left[ \left( g^{-1}\,{\rm d}g\right)^3\right] \right)\,,
\end{equation}
where the sum runs over all simple and abelian factors of $\mathcal G^{\rm up}$, $I=\{{\rm sl},\,{\rm su},\,t,\,y\}$. In our conventions we take ${\rm sgn}_I$ positive for ${ SL}(2,\mathbb R)$ and negative for the other factors.

The bosonic sector of the theory on $\mathcal G^{\rm up}$ has levels $\mathtt k_{\rm sl}= n_5 + 2$, $\mathtt k_{\rm su} = n_5 -2$, $\mathtt k_t =2$ and $\mathtt k_y = 2 R_y^2$, while in the full supersymmetric theory additional fermionic contributions set these to 
  $\mathtt k_{\rm sl}= \mathtt k_{\rm su}=n_5$. As we are primarily interested in backgrounds within the supergravity regime $n_5 \gg 1$, we will adopt the latter values throughout. Explicitly we find the upstairs model to be:\footnote{An ambiguity in the choice of integration constant of the two-form $B$-field has been discussed in previous literature (see e.g.~\cite{Martinec:2023plo} and references therein). Since it does not affect our analysis, we shall not elaborate on this here.}
\begin{equation}
\label{eq:upWZWmodel}
\begin{aligned}
S_{WZW} &= \frac{n_5}{\pi} \int_{\Sigma_2}{\rm d}^2z\,\Big[ \partial \rho\, \overline{\partial} \rho + \cosh^2\rho\,\partial \sigma \,\overline\partial \sigma- \sinh^2\rho\,\partial\tau\,\overline\partial\tau \\
&+ \partial\theta\,\overline\partial\theta + \sin^2\theta \,\partial\phi \,\overline \partial\phi + \cos^2\theta\, \partial\psi \,\overline \partial\psi 
\Big]+ \frac{1}{\pi}\int_{\Sigma_2}{\rm d}^2z\Big[ \partial y \,\overline \partial y - \partial t\,\overline \partial t\Big]  \\
&+ \frac{n_5}{\pi} \int_{\Sigma_2}{\rm d}^2z\,\Big[ \sinh^2\rho \left(\partial \sigma\, \overline\partial \tau - \partial \tau \,\overline \partial \sigma \right) - \cos^2\theta \left(\partial \phi\, \overline \partial \psi - \overline \partial\phi\, \partial \psi \right)\Big]\,.
\end{aligned}
\end{equation} 
The target space metric and $H$-field on the eight-dimensional group manifold then take the form
\begin{equation}
\label{eq:upmetric}
\begin{aligned}
\diff s_{\rm up}^2 &= n_5 \Bigl[ -\sinh^2\rho \,{\rm d}\tau^2 + {\rm d}\rho^2 +\cosh^2\rho\,{\rm d}\sigma^2 + {\rm d}\theta^2 + \sin^2 \theta \,{\rm d}\phi^2 + \cos^2\theta\,{\rm d}\psi^2\Bigr]
- {\rm d}t^2 + {\rm d}y^2,\\
H_3 &= n_5\Bigl[ \sinh2\rho\,\diff \rho \wedge \diff \sigma\wedge \diff \tau + \sin 2\theta \,\diff \theta \wedge \diff \phi\wedge \diff \psi\Bigr]\,.
\end{aligned}
\end{equation}

The upstairs model \eqref{eq:upWZWmodel} has a set of sixteen conserved holomorphic and antiholomorphic currents. We introduce a convenient set of generators for ${SL}(2,\mathbb R)$ and ${SU}(2)$. For the former, we take
\begin{equation}
\label{eq:genSL}
t^1_{\rm sl} = -\frac{1}{2}\sigma_2\,,\qquad t^2_{\rm sl} = \frac{i}{2}\sigma_1\,,\qquad t^3_{\rm sl} = \frac{i}{2}\sigma_3\,,
\end{equation}
while for ${SU}(2)$ we introduce
\begin{equation}
\label{eq:genSU}
\begin{aligned}
t^1_{\rm su}&= \frac{1}{2}\sigma_1\,,\qquad t^2_{\rm su} = \frac{1}{2}\sigma_2\,,\qquad  t^3_{\rm su} = \frac{1}{2}\sigma_3\,,
\end{aligned}
\end{equation}
The six conserved worldsheet currents associated to the ${\rm sl}(2,\mathbb R)$ algebra can, then, be expressed as 
\begin{equation}
\label{eq:sl2currents}
\begin{aligned}
\mathcal J^\pm_{\rm sl} &= -i \,\kk_{\rm sl}\,{\rm Tr}\left[ \left( t^1_{\rm sl} \pm t^2_{\rm sl} \right) \,\partial g_{\rm sl}\,g^{-1}_{\rm sl}\right] = {\rm e}^{\mp\left( \tau + \sigma\right)}n_5\left[\pm \partial\rho -\frac{1}{2}\sinh2\rho\left( \partial\tau - \partial\sigma\right) \right]\,,\\[1mm]
\mathcal J^3_{\rm sl} &= -i\, \kk_{\rm sl}\,{\rm Tr}\left[ t^3_{\rm sl} \,\partial g_{\rm sl} \,g^{-1}_{\rm sl} \right] = n_5 \left[ \cosh^2\rho\, \partial\sigma - \sinh^2\rho\, \partial\tau\right]\,,\\[1mm]
\bar {\mathcal J}^\pm_{\rm sl} &= -i\,\kk_{\rm sl}\, {\rm Tr}\left[ \left( t^1_{\rm sl} \pm t^2_{\rm sl} \right) \,g^{-1}_{\rm sl}\,\bar \partial g_{\rm sl}\right] = {\rm e}^{\mp\left( \tau - \sigma\right)}n_5\left[\pm \bar\partial\rho - \frac{1}{2}\sinh2\rho\left( \bar\partial\tau + \bar\partial\sigma\right) \right]\,,\\[1mm]
\bar {\mathcal J}^3_{\rm sl} &=-i \,\kk_{\rm sl}\, {\rm Tr}\left[ t^3_{\rm sl} \,g^{-1}_{\rm sl}\,\bar \partial g_{\rm sl} \right] = n_5 \left[ \cosh^2\rho 
\,\bar\partial\sigma + \sinh^2\rho \,\bar\partial\tau\right]\,,
\end{aligned}
\end{equation}
and analogous expressions for the six conserved currents generating the ${\rm su}(2)$ algebra. In particular, we will need in the following the explicit form of $\mathcal J^3_{\rm su}$, $\bar{\mathcal J}^3_{\rm su}$:
\begin{equation}
\label{eq:su2currents}
\begin{aligned}
\mathcal J^3_{\rm su} &= n_5\left( \cos^2\theta\,\partial\psi - \sin^2\theta\, \partial\phi\right)\,,\\[1mm]
\bar{\mathcal J}^3_{\rm su} &= n_5\left( \cos^2\theta\,\bar \partial\psi + \sin^2\theta\, \bar \partial\phi\right)\,.
\end{aligned}
\end{equation}
Finally, the currents along the timelike and spacelike factors $\mathbb R\times {\rm U}(1)$ are denoted as
\begin{equation}
\label{eq:tymomenta}
{\cal P}_t = \partial t\,\qquad \bar{\cal P}_t = \bar\partial t\,,\qquad {\cal P}_y = \partial y\,,\qquad \bar {\cal P}_y = \bar\partial y\,.
\end{equation}


\subsection{The gauged WZW model}
\label{sec:downstairsmodel}

We now turn to the coset
\begin{equation}
\label{eq:cosetgroup}
\frac{SL(2,\mathbb R)\times SU(2) \times \mathbb R \times U(1)}{U(1)_L \times U(1)_R}\,.
\end{equation}
General details about the gauging procedure can be found in appendix \ref{app:WZWmodels}. The gauging is determined by specifying the chiral embeddings of the ${U}(1)$ factors into the left and right isometry groups of the upstairs group manifold, $\varphi_L : U(1)_L \rightarrow \mathcal G^{\rm up}_L$ and $\varphi_R: U(1)_R \rightarrow \mathcal G^{\rm up}_R$. These embeddings define the group action being gauged, which in our conventions takes the form 
\begin{equation}
\label{gaction}
\begin{aligned}
g\rightarrow g'=\Big( g'_{\rm sl}\,,\,g'_{\rm su}\,,\,e^{t+ l_3 h_L + r_3 h_R}\, ,\,e^{i \frac{y- l_4 h_L - r_4 h_R}{R_y}} \Big),
\end{aligned}
\end{equation}
with 
\begin{equation}
\begin{aligned}
g'_{\rm sl} &= e^{\frac{1}{2}(\tau + \sigma- 2l_1 h_L)\sigma_3}e^{\rho\,\sigma_1}e^{\frac{1}{2}(\sigma-\tau-2 r_1 h_R)\sigma_3}\,,\\
g'_{\rm su} &= e^{\frac{i}{2}(\psi - \phi-2 l_2 h_L)\sigma_3}e^{i\theta\,\sigma_1}e^{\frac{i}{2}(\psi + \phi- 2r_2 h_R)\sigma_3}\,,
\end{aligned}
\end{equation}
corresponding to the chiral group embeddings
\begin{equation}
\begin{aligned}
\varphi_L(h_L) &= \left(- l_1 h_L \sigma_3\,,\,-i l_2 h_L \sigma_3\,,\,l_3 h_L\,,\,- i \frac{l_4}{R_y}h_L\right)\,, \qquad \varphi_R(h_L) = 0\,,\\
\varphi_R(h_R) &= \left( r_1 h_R \sigma_3\,,\,i r_2 h_R \sigma_3\,,\,-r_3 h_R\,,\, i \frac{r_4}{R_y}h_R\right)\,, \qquad \varphi_L(h_R) =0\,,
\end{aligned}
\end{equation}
for some real $h_L$ and $h_R$, and real gauging parameters $l_n$ and $r_n$, $n=1,2,3,4$. The resulting gauge transformations act on the coordinates as
\begin{equation}
\label{eq:gaugefreedom}
\begin{aligned}
\delta\tau&= r_1 h_R - l_1 h_L \,,\qquad \qquad \delta\sigma = -\left( l_1 h_L + r_1 h_R\right)\,,\\
\delta\phi &=l_2 h_L - r_2 h_R\,,\qquad \qquad \delta\psi = -\left( l_2 h_L + r_2 h_R\right)\,,\\
\delta t &=l_3 h_L + r_3 h_R \,,\qquad \qquad \delta y = -\left(l_4 h_L + r_4 h_R\right)\,.
\end{aligned}
\end{equation}
The action for the gauged WZW model is given by \eqref{eq:gaugedsigma} and \eqref{eq:generalgWZWmodel}. One introduces target-space one-forms $\theta_{1,2}$ given by
\begin{equation}
\begin{aligned}
\theta_1 \,&=\,n_5\left( -l_1 \text{Tr}\left[ \frac{\sigma_3}{2}\, {\rm d}g_{\rm sl}\,g_{\rm sl}^{-1}\right] + l_2 \text{Tr}\left[ i\frac{\sigma_3}{2}\, {\rm d}g_{\rm su}\,g_{\rm su}^{-1}\right]\right) - l_3 {\rm d}t - l_4 {\rm d}y\\[2mm]
\,&=\,  n_5 \Bigl(l_1 \left( \sinh^2\rho\, {\rm d}\tau - \cosh^2\rho\, {\rm d}\sigma\right) + l_2 \left(\sin^2\theta {\rm d}\phi- \cos^2\theta {\rm d}\psi \right) \Bigr)- l_3 {\rm d}t - l_4 {\rm d}y\,,
\end{aligned}
\end{equation}
and
\begin{equation}
\begin{aligned}
\theta_2 \,&=\, n_5\left(r_1 \text{Tr}\left[ \frac{\sigma_3}{2}\, g_{\rm sl}^{-1}\,{\rm d}g_{\rm sl}\right] - r_2 \text{Tr}\left[ i\frac{\sigma_3}{2}\, g_{\rm su}^{-1}\,{\rm d}g_{\rm su}\right]\right) + r_3 {\rm d}t + r_4 {\rm d}y\\[2mm]
\,&=\,  n_5 \Bigl(r_1 \left( \sinh^2\rho\, {\rm d}\tau + \cosh^2\rho\, {\rm d}\sigma\right) +r _2 \left( \sin^2\theta {\rm d}\phi + \cos^2\theta {\rm d}\psi\right) \Bigr)+ r_3 {\rm d}t + r_4 {\rm d}y\,.
\end{aligned}
\end{equation}
Each one-form $\theta_{1,2}$ corresponds to a dual Killing vector $\xi_{1,2}$ associated to the group action \eqref{gaction} being gauged. Explicitly:
\begin{equation}
\label{eq:gaugingKV}
\begin{aligned}
\xi_1 &\,=\, -l_1 \left( \partial_\tau + \partial_\sigma\right) + l_2 \left( \partial_\phi - \partial_\psi\right) + l_3 \partial_t - l_4 \partial_y\,,\\[1mm]
\xi_2 &\,=\, r_1 \left( \partial_\tau - \partial_\sigma\right) - r_2 \left( \partial_\psi + \partial_\phi\right) + r_3 \partial_t - r_4 \partial_y\,.
\end{aligned}
\end{equation}
The gauged model is anomaly-free if we gauge a null subgroup of isometries~\cite{Klimcik:1994wp,Israel:2004ir, Martinec:2017ztd}, which requires the Killing vectors to be null with respect to the metric \eqref{eq:upmetric}. This leads to the null constraints:
\begin{equation}
\label{eq:nullconstraint}
\begin{aligned}
n_5\left( l_1^2 + l_2^2 \right) - l_3^2 + l_4^2=0\,, \qquad \qquad n_5\left( r_1^2 + r_2^2 \right) - r_3^2 + r_4^2=0\,.
\end{aligned}
\end{equation} 
By pulling back the one-forms $\theta_{1,2}$ we introduce worldsheet currents $\mathcal J$, $\bar {\mathcal J}$. These can be written in terms of the fundamental group currents \eqref{eq:sl2currents}, \eqref{eq:su2currents}, \eqref{eq:tymomenta} as
\begin{equation}
\begin{aligned}
\label{eq:gauging_currents}
\mathcal J &=  l_1 {\cal J}_{\rm sl}^3 + l_2 {\cal J}_{\rm su}^3 + l_3 {\cal P}_t+ l_4 {\cal P}_y=\\[2mm]
&= -n_5\Bigl[ l_1\left( \sinh^2\rho\,\partial\tau - \cosh^2\rho \,\partial\sigma\right) + l_2 \left( \sin^2 \theta\, \partial\phi - \cos^2\theta\, \partial\psi \right)\Bigr]+ l_3\, {\partial}t + l_4 \,{\partial}y\,,\\[1mm]
\bar{\mathcal J} &=  r_1 \bar{{\cal J}}_{\rm sl}^3 + r_2 \bar{{\cal J}}_{\rm su}^3 + r_3 \bar {\cal P}_t+ r_4 \bar {\cal P}_y=\\[2mm]
&= n_5\Bigl[ r_1\left( \sinh^2\rho\,\bar\partial\tau + \cosh^2\rho\, \bar\partial\sigma\right) + r_2 \left( \sin^2\theta\,\bar \partial\phi + \cos^2 \theta \,\bar\partial\psi\right)\Bigr]+ r_3\, \bar{\partial}t + r_4 \,\bar{\partial}y\,.
\end{aligned}
\end{equation}
The resulting gauged WZW model is then specified by the action 
\begin{equation}
\label{eq:nonextrgWZWmodel}
S_{gWZW} = S_{WZW} +\frac{2}{\pi} \int_{\Sigma_2}{\rm d}^2z\big[\mathcal A\, \bar{{\cal J}} +  \bar{\mathcal A}\,\mathcal{J} - 2\Sigma\, \mathcal A\, \bar {\mathcal A}\big]\,,
\end{equation}
where the upstairs theory $S_{WZW}$ is given by \eqref{eq:upWZWmodel} and
\begin{equation}\label{eq:sigma_function}
\Sigma = -\frac{1}{2}\xi^M_1 \,g_{MN}\, \xi^N_2=-\frac{1}{2}\Bigl[ n_5 \left( l_1 r_1 \cosh2\rho + l_2 r_2 \cos2\theta\right) + \left( l_4 r_4 - l_3 r_3\right) \Bigr]\,,
\end{equation}
where $g_{MN}$ is the upstairs metric \eqref{eq:upmetric}.
The worldsheet gauge fields enter the Lagrangian only quadratically and can therefore be integrated out. The resulting sigma model can be directly compared with the corresponding supergravity background in the large-level limit. 
After integrating out the gauge fields we obtain
\begin{equation}
\label{eq:effectivegWZW}
S_{gWZW} = S_{WZW} +\frac{1}{\pi} \int_{\Sigma_2}{\rm d}^2z\,\Sigma^{-1}\,\mathcal J \,\bar{\mathcal J}\,.
\end{equation}
In the next section we present the background fields obtained from this action and show that these can be matched to the supergravity solution \eqref{eq:decoupledmetric}.


\subsection{Supergravity fields}

In the previous sections we constructed a gauged WZW model that provides a worldsheet description of string propagation on the coset~\eqref{eq:cosetgroup}. 
In this section we analyze the metric, dilaton, and B-field arising from the gauged model \eqref{eq:effectivegWZW} and determine the values for the real coefficients $l_n$, $r_n$ (with $n=1,2,3,4$) for which the target space precisely matches the supergravity background \eqref{eq:decoupledmetric}. Our focus is on the black hole regime of the solution discussed in section \ref{sec:decoupling}, where we take $0\leq \ell<1$ and $b<M$. However, we will also briefly comment on the horizonless regime. Indeed, the background obtained here is related to the horizonless soliton analyzed in~\cite{Bufalini:2021ndn} through an analytic continuation, as we will discuss shortly.  

To express the background in a convenient form, we fix the gauge freedom \eqref{eq:gaugefreedom} by setting $\sigma=\tau=0$.
The six-dimensional metric then takes the form
\begin{equation}
\label{eq:backgroundtargetspace}
\begin{aligned}
{\rm d}s^2 &= \Sigma^{-1}\left( -h_{tt} \,{\rm d}t^2+ h_{yy} \,{\rm d}y^2\right) + \left(l_4 r_3 + l_3 r_4 \right)\Sigma^{-1}\,{\rm d}t \,{\rm d}y\\[1mm]
&\quad + n_5 \left( {\rm d}\rho^2 + {\rm d}\theta^2\right)+n_5\,\Sigma^{-1}\Bigl[ h_{\phi\phi} \,\sin^2\theta {\rm d}\phi^2+ h_{\psi\psi} \,\cos^2\theta {\rm d}\psi^2\Bigr] \\[1mm]
&\quad +n_5\,\Sigma^{-1}\,\sin^2\theta\Bigl[ \left(l_3 r_2 - l_2 r_3\right) {\rm d}t + \left( l_4 r_2 - l_2 r_4\right) {\rm d}y\Bigr]{\rm d}\phi\\[1mm]
&\quad +n_5\,\Sigma^{-1}\, \cos^2\theta \Bigl[ \left(l_3 r_2 + l_2 r_3\right) {\rm d}t + \left(l_4 r_2 + l_2 r_4\right) {\rm d}y\Bigr]{\rm d}\psi \,,
\end{aligned}
\end{equation} 
with
\begin{equation}
\begin{aligned}
h_{tt} &= -\frac{1}{2}\Bigl[n_5 \left( l_2 r_2 \cos2\theta + l_1 r_1 \cosh2\rho\right) + l_3 r_3 + l_4 r_4\Bigr]\,,\\[1mm]
h_{yy} &= \frac{1}{2}\Bigl[-n_5 \left( l_2 r_2 \cos2\theta + l_1 r_1 \cosh2\rho\right) + l_3 r_3 + l_4 r_4\Bigr]\,,\\[1mm]
h_{\phi\phi} &= \frac{1}{2}\Bigl[-n_5\left(l_2 r_2 + l_1 r_1 \cosh2\rho \right)+ l_3 r_3 - l_4 r_4\Bigr]\,,\\[1mm]
h_{\psi\psi} &= \frac{1}{2}\Bigl[ n_5\left(l_2 r_2 - l_1 r_1 \cosh2\rho \right)+ l_3 r_3 - l_4 r_4\Bigr]\,.
\end{aligned}
\end{equation}
The associated B-field reads
\begin{equation}
\begin{aligned}
B &= \frac{l_3r_4-l_4 r_3}{2\Sigma}\diff t\wedge \diff y + \frac{n_5\,h_{\phi\phi}}{\Sigma}\cos^2\theta \diff\phi \wedge \diff \psi \\
&\quad + \frac{1}{2\Sigma}\Bigl[  \left(l_3 r_2 + l_2 r_3\right) {\rm d}t+ \left(l_4 r_2 + l_2 r_4\right) {\rm d}y \Bigr]\wedge \sin^2\theta \diff \phi \\
&\quad + \frac{1}{2\Sigma}\Bigl[ \left(l_3 r_2 - l_2 r_3\right) {\rm d}t + \left( l_4 r_2 - l_2 r_4\right) {\rm d}y\Bigr] \wedge \cos^2\theta \diff \psi\,.
\end{aligned}
\end{equation}
The dilaton field determined by solving the supergravity field equations on such background is of the form ${\rm e}^{2\Phi} \sim n_5/\Sigma$, in accordance with \eqref{eq:decoupledmetric}.  

The metric \eqref{eq:backgroundtargetspace} is mapped to the supergravity black hole given by \eqref{eq:decoupledmetric} by setting $q_5 = n_5$ (we are setting $\alpha' =1$) and choosing the embedding coefficients as follows:
\begin{equation}
\label{eq:gaugingparameterssugra}
\begin{aligned}
r_1 &= - 8\sqrt{\frac{q_1\,q_p\left( q_1 + q_p\right)\left( M^2 - b^2\right)}{q_5\left(M^2 + 16 \,q_1\, q_p\right)}}\,,\qquad r_2 = 8\, b\,\sqrt{\frac{q_1\,q_p\left( q_1 + q_p\right)}{q_5\left(M^2 + 16\, q_1\, q_p\right)}}\,,\\
r_3 &= \sqrt{\left(q_1 + q_p\right) \left( M^2 + 16 \,q_1 \,q_p\right)}\,,\qquad r_4 = \left(M^2 - 16\, q_1 \,q_p\right) \sqrt{\frac{q_1+q_p}{M^2 + 16\, q_1\,q_p}}\,,\\
l_1&= 2\sqrt{\frac{q_1\,q_p\left( M^2 + 16\,q_1\,q_p\right) \left( 1- \ell^2\right)}{q_5\left( q_1 + q_p\right)}}\,,\qquad l_2 = 2\ell\sqrt{\frac{q_1\,q_p\left(M^2 + 16 \,q_1\,q_p\right)}{q_5\left( q_1 + q_p\right)}}\,,\\
l_3 &= \sqrt{\left(q_1 + q_p\right) \left( M^2 + 16 \,q_1 \,q_p\right)}\,,\qquad l_4 = \left( q_1 - q_p\right) \sqrt{\frac{M^2 + 16 \,q_1\,q_p}{q_1+ q_p}}\,.
\end{aligned}
\end{equation}
This parametrization is chosen to automatically satisfy both null constraints \eqref{eq:nullconstraint}, as well as the relations
\begin{equation}
\label{eq:cond1}
l_3 = r_3 \,,\qquad \qquad r_1 \,l_1 <0\,.
\end{equation}
We will encounter these two conditions again in the following discussion, particularly when analyzing the absence of closed timelike curves in the target space metric and the consistency of the worldsheet spectrum. 
We also observe that the coefficients \eqref{eq:gaugingparameterssugra} can be expressed as
\begin{equation}\label{eq:ratio_charges}
    \frac{r_4}{l_4}= \frac{Q_1 + Q_p}{Q_p-Q_1}\,,\qquad \qquad \frac{r_2}{l_2}= \frac{J_+}{J_-}\,,
\end{equation}
in terms of the background charges \eqref{eq:bhcharges}.

\medskip

In the regime where the supergravity background admits a horizon, i.e. $0\leq \ell <1$ and $b<M$, the coefficients $l_1$ and $r_1$ always remain real. However, in the opposite regime, corresponding to the branch of the solution in which the geometry is horizonless, these coefficients become imaginary. This aligns with the analysis of~\cite{Martinec:2017ztd,Martinec:2018nco,Martinec:2019wzw,Martinec:2020gkv,Bufalini:2021ndn}, where the worldsheet description of the horizonless configuration has been considered. Indeed, the background they derive  is obtained from ours by taking $l_1$ and $r_1$ to be purely imaginary, via the analytic continuation $l_1 \to -i\,\mathfrak{l}_1$ and $r_1 \to -i\, \mathfrak{r}_1$, where $\mathfrak{r}_1$ and $\mathfrak{l}_1$ are now real. In addition to this analytic continuation, obtaining a causally well-behaved horizonless soliton requires fixing one rotation parameter in terms of the other, as dictated by \eqref{eq:decsoliton}. In terms of the gauging parameters, imposing this condition leads to the simple relations
\begin{equation}
\label{eq:jmart}
\left( \mathfrak{l}_1\right)^2 = \left(\mathfrak{r}_1\right)^2\,,\qquad \qquad \mathfrak{l}_1 \mathfrak{r}_1 >0\,.
\end{equation}
In particular, it follows that $\mathfrak{l}_3/\mathfrak{l}_1 = \mathfrak{r}_3/\mathfrak{r}_1$, in agreement with~\cite{Martinec:2018nco,Bufalini:2021ndn}. Although the supergravity background obtained after this analytic continuation is real, the gauged WZW model (before integrating out the gauge fields and gauge fixing $\sigma = \tau=0$) is not. Reality is restored after taking the additional analytic continuation \eqref{eq:btztoads}, which brings us back to AdS$_3$ in global coordinates. With both continuations applied, the resulting gauged WZW model yields a real target space metric that precisely reproduces the supergravity background of the non-supersymmetric horizonless soliton of~\cite{Jejjala:2005yu} after the NS$5$ decoupling limit.


\subsubsection{Causality and temperature}
\label{sec:causalitytemperature}

In this section we analyze in more detail the string background \eqref{eq:backgroundtargetspace}, and we demonstrate that the conditions \eqref{eq:cond1} are sufficient to ensure the absence of closed timelike curves (CTCs), whose presence would otherwise lead to causality violations. Our analysis closely follows that of~\cite{Bufalini:2021ndn}. 

To investigate the potential presence of CTCs, we rewrite the metric \eqref{eq:backgroundtargetspace} in the following adapted form:
\begin{equation}
\label{eq:adaptctcmetric}
\begin{aligned}
&\diff s^2 = - \Delta_t(\rho) \,\diff t^2 + \Upsilon(\rho)\Bigl( \diff y + \omega_y(\rho) \diff t\Bigr)^2 + n_5 \left( \diff\rho^2 + \diff \theta^2\right) \\[2mm]
&+ \frac{n_5}{\Sigma}\Bigl[ h_{\phi\phi} \sin^2\theta\Bigl( \diff \phi + \omega_\phi(\rho) \diff t + \chi_\phi(\rho) \diff y\Bigr)^2 + h_{\psi\psi}\cos^2\theta\Bigl( \diff\psi + \omega_\psi(\rho) \diff t + \chi_\psi(\rho) \diff y\Bigr)^2\Bigr].
\end{aligned}
\end{equation}
To keep the presentation lighter, we do not explicitly report the expressions for the functions $\Delta_t$, $\Upsilon$, $\omega_y$, $\omega_\phi$, $\omega_\psi$, $\chi_\phi$ and $\chi_\psi$ as they can be readily obtained by comparing \eqref{eq:adaptctcmetric} with \eqref{eq:backgroundtargetspace}. Note that the function $\Delta_t$ vanishes at $\rho=0$ by imposing  \eqref{eq:nullconstraint}, indicating the presence of an event horizon at that location.

The absence of CTCs outside and on the horizon requires the following conditions to hold:
\begin{equation}
\label{eq:absencectc}
\Sigma^{-1}h_{\phi\phi}\geq 0\,,\qquad \Sigma^{-1}h_{\psi\psi} \geq 0,\,\qquad \Upsilon \geq 0\,,
\end{equation}
for $\rho \geq 0$. To verify that these conditions are satisfied, we first rewrite the function $\Sigma$ \eqref{eq:sigma_function} by imposing $l_3 = r_3$ and using the null constraints \eqref{eq:nullconstraint} to eliminate the dependence on $r_3^2$, obtaining
\begin{equation}
4\Sigma= \left( l_4 - r_4\right)^2 + n_5 \left( l_1 + r_1 \right)^2 + n_5 \left( l_2 \mp r_2\right)^2 + 2 n_5\Bigl[ l_2 r_2 \left(\pm 1- \cos2\theta\right) - l_1 r_1\left( 1 + \cosh2\rho\right)\Bigr].
\end{equation}
The minimum of $\Sigma$ depends on the sign of the product $l_2 r_2$:
\begin{equation}
\Sigma_{\rm min} = \begin{cases}
\Sigma\left( \rho ,\theta =0\right) \,,& {\rm if} \,\,l_2 r_2 \geq 0\,,\\
\Sigma\left( \rho ,\theta = \pi/2\right)\,,& {\rm if} \,\,l_2 r_2 <0\,.
\end{cases}
\end{equation}
In both cases the following bound holds
\begin{equation}
4\Sigma >  \left( l_4 - r_4\right)^2 + n_5 \left( l_1 + r_1 \right)^2 + n_5 \left( l_2 \mp r_2\right)^2 -2 n_5\, l_1 r_1\left( 1 + \cosh2\rho\right)\,.
\end{equation}
Since the last term is always positive when $l_1 r_1 <0$, it follows that $\Sigma>0$ for all values of $\theta$ and $\rho$. Moreover, since $h_{\phi\phi} = \Sigma(\rho,\theta=0)$ and $h_{\psi\psi} = \Sigma(\rho,\theta = \pi/2)$, these functions are also positive if $l_1 r_1 <0$, thereby satisfying the first two conditions in \eqref{eq:absencectc}. 

We now turn to the function $\Upsilon$, which can be shown to be
\begin{equation}
\begin{aligned}
4 \,h_{\phi\phi}\,h_{\psi\psi}\,\Upsilon &= \left(n_5\, l_1^2 - l_3^2\right)\left( n_5\, r_1^2 - r_3^2\right) - \left( n_5\, l_2^2 + l_4^2\right) \left( n_5\, r_2^2 + r_4^2 \right) + n_5\left( l_1 r_3 + l_3 r_1\right)^2 \\[2mm]
& + n_5^2\, l_1^2 r_1^2 \sinh^22\rho - 2 n_5 \,l_1 r_1 l_3 r_3 \left( 1+ \cosh2\rho\right)\,.
\end{aligned}\end{equation}
 Using the null constraints \eqref{eq:nullconstraint} the first two terms in the sum cancel, and the remaining ones are manifestly positive when $l_1 r_1 <0$ and $l_3 = r_3$. Thus, we have established that the conditions \eqref{eq:cond1} imply the validity of \eqref{eq:absencectc}, ensuring the absence of closed timelike curves on the background \eqref{eq:backgroundtargetspace}. 
 
We now turn to the computation of the inverse Hawking temperature associated to the background, following the standard approach of~\cite{Gibbons:1976ue}. To do so, we express the metric in an adapted form:
\begin{equation}
 \label{eq:adaptedtempmetric}
 \begin{aligned}
 \diff s^2 &= - f_{tt}\diff t^2 + n_5 \left( \diff \rho^2 + \diff \theta^2\right) + n_5\,\Sigma^{-1}\Bigl[n_5^{-1}\,h_{yy} \left( \diff y - \omega_y \diff t\right)^2 + h_{\phi\phi}\sin^2\theta \left( \diff \phi - \omega_\phi \diff t\right)^2
 \\[2mm]
 &+h_{\psi\psi}\cos^2\theta\left(\diff\psi - \omega_\psi \diff t\right)^2 + \left( l_4 r_2-l_2 r_4\right)\sin^2\theta \left( \diff y - \omega_y \diff t\right) \left( \diff \phi - \omega_\phi \diff t\right)\\[2mm]
&+ \left( l_4 r_2 + l_2 r_4\right) \cos^2\theta \left( \diff y - \omega_y \diff t\right)\left(\diff\psi - \omega_\psi \diff t\right)\Bigr]\,,
 \end{aligned}
 \end{equation}
 where $f_{tt},\,\omega_\phi,\,\omega_\psi,\,\omega_y$ are functions that can be determined by comparing this ansatz with \eqref{eq:backgroundtargetspace}. The presence of an event horizon is indicated by the vanishing of $f_{tt}$ at $\rho = 0$,
 \begin{equation}
\lim_{\rho\rightarrow 0} f_{tt} =0\,.
 \end{equation}
 We define the angular velocities of the horizon as the limiting values of the functions $\omega_{\phi,\psi,y}$:
 \begin{equation}
 \Omega_{\phi,\psi,y} \equiv \lim_{\rho\rightarrow 0} \omega_{\phi,\psi,y}\,.
 \end{equation}
 Their explicit expressions are given by
 \begin{equation}
 \label{eq:angular_velocity}
 \Omega_\phi = -\frac{l_1r_2 + l_2 r_1}{l_3\left(l_1 -r_1\right)}\,,\qquad \Omega_\psi =  -\frac{l_1r_2 - l_2 r_1}{l_3\left(l_1 -r_1\right)} \,,\qquad \Omega_y =  -\frac{l_1r_4 - l_4 r_1}{l_3\left(l_1 -r_1\right)}  \,.
 \end{equation}
This rewriting of the metric makes it clear that the event horizon is generated by the Killing vector
 \begin{equation}
 \xi = \partial_t + \Omega_\phi \partial_\phi + \Omega_\psi \partial_\psi + \Omega_y \partial_y\,,
 \end{equation}
 whose norm vanishes at $\rho =0$. The one-forms $\omega_y\, \diff t$, $\omega_\phi\,\diff t$ and $\omega_\psi\,\diff t$, describing the fibration of the compact internal ${S}^3 \times {S}^1$ over the time, are not well-defined on the horizon. To ensure regularity, we introduce new angular coordinates:
 \begin{equation}
 \tilde\phi = \phi - \Omega_\phi \,t \,\qquad \tilde\psi = \psi - \Omega_\psi \,t\,\qquad \tilde y = y - \Omega_y\, t\,,
 \end{equation}
with standard periodicity, $\tilde\phi \sim \tilde\phi + 2\pi$, $\tilde\psi \sim \tilde\psi + 2\pi$ and $\tilde y \sim \tilde y + 2\pi R_y$. To compute the Hawking temperature, we analytically continue the metric to Euclidean signature by performing the Wick rotation $t\rightarrow -i\tau$. The Euclidean time $\tau$ is compactified with period $\beta$, i.e. $\tau \sim \tau + \beta$. The inverse Hawking temperature $\beta$ is then determined by requiring that the Euclidean solution caps off smoothly at $\rho=0$, without conical singularities. Expanding the relevant part of the metric around $\rho \sim 0$, we find
\begin{equation}
\diff s^2 = n_5\left( \diff \rho^2 + \rho^2 \left( \frac{2\pi}{\beta}\right)^2 \diff \tau^2 \right) + ... \,,
\end{equation}
with
\begin{equation}
\label{eq:beta}
\beta = -\frac{\pi \,l_3}{l_1\,r_1}\left( l_1 - r_1\right)\,.
\end{equation}
This confirms that the background describes a finite-temperature rotating black hole endowed with a Killing horizon.

\subsubsection{Interesting limits}

\paragraph{JMaRT.}
As argued above, the regime in which the background represents a horizonless soliton can be obtained by analytically continuing the parameters $l_1 \to -i \,\mathfrak{l}_1$ and $r_1 \to -i \,\mathfrak{r}_1$, and then imposing the condition \eqref{eq:jmart}, which translates into $\mathfrak{l}_1 = \mathfrak{r}_1$. This results in a vanishing inverse temperature \eqref{eq:beta}, $\beta \to 0$. Furthermore, in this limit the angular velocities \eqref{eq:angular_velocity} diverge, but the following combinations remain finite,
\begin{equation}
    \beta \Omega_\phi = i \pi \left( \mathtt l_2 + \mathtt r_2\right)\,,\qquad \beta \Omega_\psi = -i\pi \left( \mathtt l_2 - \mathtt r_2\right)\,,\qquad \beta \Omega_y = -i\pi \left( \mathtt l_4 - \mathtt r_4\right)\,,
\end{equation}
where we introduced the new parameters $\mathtt r_n = \frac{r_n}{\mathfrak{r}_1}$ and $\mathtt l_n = \frac{l_n}{\mathfrak{l}_1}$, for $n = 2,3,4$. It is clear that in this limit the Killing vector whose orbits contract at $\rho \to 0$ is the spacelike combination given by
\begin{equation}
    \xi_{\rm JMaRT} = \partial_y + \frac{\Omega_\phi}{\Omega_y}\partial_{\phi} + \frac{\Omega_\psi}{\Omega_y}\partial_\psi\,.
\end{equation}
Examining the global properties of the solution in this regime, one finds that the geometry caps off smoothly at $\rho \to 0$, where the vector $\xi_{\rm JMaRT}$ shrinks, provided a specific relation between the gauging parameters and the radius $R_y$ is satisfied. To determine this relation, we expand the metric near $\rho \sim 0$, introducing the angular coordinates $\tilde\phi = \phi - \frac{\Omega_\phi}{\Omega_y}y$ and $\tilde\psi = \psi - \frac{\Omega_\psi}{\Omega_y}y$, which yields 
\begin{equation}
    \diff s^2 = n_5 \Bigl[ \diff \rho^2 + \rho^2 \left(\frac{2\diff y}{\mathtt r_4 - \mathtt l_4}\right)^2\Bigr] + \,.\,.\,.\,.
\end{equation}
The geometry caps off smoothly provided\footnote{One may also allow for simple orbifolds, though we avoid this here for simplicity of notation.}
\begin{equation}
    R_y = \frac{\mathtt r_4-\mathtt l_4}{2}\,.
\end{equation}
In addition, the shifted angles $\tilde\phi$ and $\tilde\psi$ have the standard $2\pi$ periodicity only if
\begin{equation}
\begin{aligned}
    R_y\frac{\Omega_\phi}{\Omega_y} = R_y\frac{\mathtt l_2 + \mathtt r_2}{\mathtt r_4 - \mathtt l_4} \in\mathbb Z \,\implies \, \mathtt l_2 + \mathtt r_2\in 2\mathbb Z\,,
    \\[1mm]
    R_y\frac{\Omega_\psi}{\Omega_y} = R_y\frac{\mathtt r_2 - \mathtt l_2}{\mathtt r_4 - \mathtt l_4} \in\mathbb Z \,\implies \, \mathtt r_2 - \mathtt l_2\in 2\mathbb Z\,,
    \end{aligned}
\end{equation}
which reproduce precisely the quantization conditions for the horizonless soliton of~\cite{Jejjala:2005yu}, in agreement with the analysis of~\cite{Bufalini:2021ndn}.\footnote{Our parametrization can be mapped to that of secs.~3.2–3.3 in\cite{Martinec:2018nco} (with orbifold parameter taken as $k=1$) via
\begin{equation}
\begin{aligned}
2R_y&= \mathtt l_4- \mathtt r_4 \,,\qquad \vartheta = \frac{Q_p}{Q_1} = \frac{\mathtt r_4+\mathtt l_4}{\mathtt r_4 - \mathtt l_4}\,, \qquad R_y\varrho = -\mathtt l_3= - \mathtt r_3\,,\qquad \mathbf m + \mathbf n= \mathtt l_2\,,\qquad \mathbf m- \mathbf n = \mathtt r_2\,.
\end{aligned}
\end{equation}} 
It is clear that the limit $\beta \to 0$ corresponds to a topology change in which the horizon disappears and the geometry caps off (smoothly or up to possible orbifold singularities).

A closely related phenomenon was recently observed in~\cite{Cassani:2024kjn} in a different setting, namely in the study of asymptotically flat saddles of the five-dimensional Euclidean gravitational path integral under supersymmetric boundary conditions. There, it was shown that such Euclidean saddles, which are supersymmetric yet ``non-extremal'' (meaning that they have a finite inverse Hawking temperature $\beta$), interpolate between the supersymmetric and extremal five-dimensional black hole of~\cite{Breckenridge:1996is} and the two-center supersymmetric microstate geometry~\cite{Bena:2004de,Bena:2005va,Bena:2007kg}, related to the JMaRT solution by imposing supersymmetry and reducing along $S^1\times T^4$. The interpolation requires implementing a suitable analytic continuation, as well as a limit of the inverse temperature: the supersymmetric and extremal black hole is obtained in the $\beta \to +\infty$ limit of the finite-$\beta$ saddle, whereas the opposite $\beta \to 0$ limit leads to the horizonless configuration. More details can be found in~\cite{Cassani:2024kjn} (see also~\cite{Cassani:2025iix} for related examples involving multi-center microstate geometries). 
Thus, our analysis shows that the solitonic configuration is recovered by taking the $\beta \to 0$ limit, together with an analytic continuation of certain parameters, also in non-supersymmetric backgrounds. 

This analytic continuation from the black hole to the horizonless regime is realized on the worldsheet by passing from a model in which we gauge the hyperbolic generator of $SL(2,\mathbb R)$ to one in which the elliptic one is gauged. This explicit construction, then, offers a window to investigate the intriguing connection between black holes and horizonless solitons, by comparing the string spectrum in the $\beta \to 0$ limit in the two regimes. 

\paragraph{Extremal limit.} The extremal limit $\beta \to +\infty$ corresponds to $l_1 r_1 \to 0$. In this regime, however, the null-gauged WZW model we constructed is no longer reliable. For instance, the worldsheet variable $\rho$, defined by Eq.~\eqref{eq:rhotor}, and introduced in the supergravity description in \eqref{eq:def_rho}, becomes ill-defined in this limit. Extremal (and supersymmetric) configurations require then a separate treatment, discussed in section~\ref{sec:susy}. As we will show there, the resolution is to adopt a different parametrization of the $SL(2,\mathbb R)$ group. This change, however, also alters the gauging procedure.


\section{Consistency of the perturbative string spectrum}
\label{sec:stringspectrum}

In the previous sections, we established that the null-gauged WZW model on the coset
\begin{equation}
\frac{{SL}(2,\mathbb R) \times {SU}(2) \times \mathbb R \times {U}(1)}{{U}(1)_L \times {U}(1)_R}\,,
\end{equation} 
provides a worldsheet description of string dynamics on the black hole background \eqref{eq:decoupledmetric}. 
In this section, we investigate properties of the spectrum of this coset CFT. The spectrum is determined by combining known results for the spectrum of propagating strings on BTZ\cite{Natsuume:1996ij,Hemming:2001we,Troost:2002wk,Hemming:2002kd,Rangamani:2007fz,Ashok:2021ffx,Martinec:2023plo} and on ${ SU}(2)$ group manifolds~\cite{Fateev:1985mm,Zamolodchikov:1986bd} (which we briefly review in sec.~\ref{sec:stringsonSLandSU}) with the formalism for gauging a subgroup. A key tool in this analysis is the formalism of worldsheet spectral flow~\cite{Maldacena:2000hw,Argurio:2000tb,Hemming:2001we,Giribet:2007wp}, which plays a central role in understanding both the ungauged and gauged models. 
%
In null-gauged models, the physical spectrum of the coset theory is constructed from the appropriate BRST operator  and it consists of a subsector of the full spectrum on the upstairs group that satisfies the null constraint.
By analyzing the gauged model in more detail we derive consistency conditions on the gauging parameters that must be satisfied in order to obtain a well-defined theory. We show that the gauging is consistent only if the independent gauging parameters can be written in terms of four integers and two continuous parameters, related to the mass and rotation parameters of the BTZ factor in the upstairs theory.

\subsection{Strings on ${ SL}(2,\mathbb R) \times { SU}(2) \times \mathbb R \times { U}(1)$}
\label{sec:stringsonSLandSU}

In this section, we review the key ingredients relevant to describing string propagation on the upstairs group manifold, neglecting the four-torus as before. We employ the parametrization introduced in~\eqref{eq:upstairsgroupelement}. As a consequence, we first focus on the WZW model on the BTZ background \eqref{eq:BTZ_metric}. String propagation on such backgrounds has been studied in detail in~\cite{Natsuume:1996ij,Hemming:2001we,Troost:2002wk,Rangamani:2007fz,Ashok:2021ffx,Martinec:2023plo}. A more complete review of string propagation in this setting is provided in appendix~\ref{app:BTZspectrum}; here, we summarize only the aspects relevant to the discussion below.

The BTZ black hole arises as an orbifold on AdS$_3$ implemented through the coordinate identifications 
\begin{equation}
\left( \tau ,\,\sigma \right) \sim \left( \tau - 2\pi \alpha_-\,,\sigma + 2\pi\alpha_+\right)\,,
\end{equation}
where $\alpha_\pm$ are real and positive parameters, satisfying $\alpha_+ > \alpha_-$. These identifications lead to the presence of a twisted sector in the theory, which can be generated via a spectral flow transformation~\cite{Hemming:2001we,Rangamani:2007fz}. The spectral flow operation acts on the current generators as
\begin{equation}
J^3_{\rm sl} \rightarrow \tilde J^3_{\rm sl} = J^3_{\rm sl} + \frac{n_5}{2}w_+\,,\qquad \bar J^3_{\rm sl} \rightarrow \tilde {\bar J}^3_{\rm sl} = \bar J^3_{\rm sl} - \frac{n_5}{2}w_-\,,
\end{equation}
where the spectral flow parameters $w_\pm$ are not necessarily equal, nor restricted to be integers. Instead, they are given in terms of the worldsheet parameters $\alpha_\pm >0$ as
\begin{equation}
w_\pm = \left( \alpha_+ \mp \alpha_- \right) \mathbf n\,,\qquad \mathbf n\in\mathbb Z\,.
\end{equation}
This shift also induces an automorphism of the Virasoro algebra: 
\begin{equation} L_n \rightarrow \tilde L_n = L_n + w_+ J^3_n + \frac{n_5}{4}w_+^2 \delta_{n,0}\,, \end{equation} 
with an analogous transformation for the antiholomorphic generators. As explained in appendix~\ref{app:BTZspectrum}, representations are labeled by the parameter $j_{\rm sl}$, which determines the value of the quadratic Casimir~\eqref{eq:sl2casimir}, and by the real eigenvalue $\lambda$ of the zero mode $J^3_0$ of the current algebra.

We now turn to the WZW model on the ${SU}(2)$ group manifold~\cite{Fateev:1985mm,Zamolodchikov:1986bd}. 
Representations of the ${\rm su}(2)$ current algebra are labeled by two quantum numbers: 
\begin{itemize}
\item the spin $j_{\rm su}\in \mathbb Z/2$, which determines the quadratic Casimir of the ${\rm su}(2)$ algebra, satisfying $0\leq j_{\rm su} \leq n_5/2$, 
\item the eigenvalue $m_{\rm su}$ of the zero mode $j_0^3$ of the current operator $J_{\rm su}^3$, whose allowed values are
\begin{equation}
m_{\rm su} = -j_{\rm su},\,-j_{\rm su} +1,\,...,\,j_{\rm su} -1 ,\,j_{\rm su}\,.
\end{equation}
\end{itemize}

Spectral flow in the ${SU}(2)$ model does not generate new physically distinct representations, but it simply reshuffles existing states. Nevertheless, it is useful to include spectral flow in the analysis, as will be clear later. The spectral flow acts as an automorphism of the current and Virasoro algebras, shifting, for instance, the current $J^3_{\rm su}$ as
\begin{equation}
J^3_{\rm su} \rightarrow \tilde J^3_{\rm su} = J^3_{\rm su} - \frac{n_5}{2}w_{\rm su} \,,\qquad \bar J^3_{\rm su} \rightarrow \tilde {\bar J}^3_{\rm su} = \bar J^3_{\rm su} - \frac{n_5}{2}\bar w_{\rm su}\,,
\end{equation}
where $w_{\rm su}$ and $\bar w_{\rm su}$ are independent integers, constrained such that $w_{\rm su} \pm \bar w_{\rm su} \in 2\mathbb{Z}$. Correspondingly, the Virasoro generators transform as
\begin{equation}
L_n \rightarrow \tilde L_n = L_n - w_{\rm su}\, j^3_n + \frac{n_5}{4}w_{\rm su}^2 \delta_{n,0}\,,
\end{equation}
with an analogous shift for the antiholomorphic sector. 

Physical states in the full ${SL}(2,\mathbb{R}) \times {SU}(2) \times \mathbb{R} \times { U}(1)$ model are further characterized by additional quantum numbers. We denote by $E$ the asymptotic energy of the state, and by $p_y$ and $\bar p_y$ the left- and right-moving momenta along the compact $y$-direction. These are expressed as 
\begin{equation} p_y = \frac{n_y}{R_y} + w_y R_y\,, \qquad \bar p_y = \frac{n_y}{R_y} - w_y R_y\,, \qquad n_y\,,w_y \in \mathbb{Z}\,, 
\end{equation} 
where $n_y$ and $w_y$ correspond, respectively, to momentum and winding along the compact circle generated by $\partial_y$.

The Virasoro constraints for the full model 
are then given by:
\begin{equation}
\label{eq:Virasoroholo}
- \frac{j_{\rm sl}\left( j_{\rm sl} +1 \right)}{n_5}+ \frac{j_{\rm su} \left( j_{\rm su} +1 \right)}{n_5}-w_+\left(\lambda - \frac{n_5}{4}w_+\right) +w_{\rm su} \left(m_{\rm su} +\frac{n_5}{4}w_{\rm su}\right) -\frac{1}{4}\left(E^2 - p_y^2\right) + N =\frac{1}{2}\,,
\end{equation}
and
\begin{equation}
\label{eq:Virasoroantiholo}
- \frac{j_{\rm sl}\left( j_{\rm sl} +1 \right)}{n_5}+ \frac{j_{\rm su} \left( j_{\rm su} +1 \right)}{n_5}+w_-\left(\bar \lambda + \frac{n_5}{4}w_-\right) +\bar w_{\rm su} \left(\bar m_{\rm su} + \frac{n_5}{4}\bar w_{\rm su}\right) -\frac{1}{4}\left(E^2 - \bar p_y^2\right) + \bar N =\frac{1}{2}\,.
\end{equation} 
Here, $N$ and $\bar N$ denote the left- and right-mover occupation numbers. 

The level matching condition can be obtained by subtracting the constraints above,
\begin{equation}
\begin{aligned}
N- \bar N &= w_- \left(\bar \lambda + \frac{n_5}{4}w_-\right) + w_+ \left(\lambda - \frac{n_5}{4}w_+\right) \\
&\quad + \bar w_{\rm su} \left(\bar m_{\rm su} + \frac{n_5}{4}\bar w_{\rm su}\right) - w_{\rm su} \left(m_{\rm su} +\frac{n_5}{4}w_{\rm su}\right)- w_y\,n_y \in \mathbb Z\,,
\end{aligned}
\end{equation}
which admits a solution provided the quantization of the momentum in ${ SL}(2,\mathbb R)$ \eqref{eq:quantcond} is imposed. 


\subsection{Physical spectrum}
\label{sec:physicalspectrum}

Physical states in the null-gauged theory are described by vertex operators that are in the cohomology of the  BRST operator constructed from the stress-tensor, the null currents and their superpartners, together with the  appropriate ghost system. This requirement imposes additional gauge constraints on the quantum numbers of physical states, to be satisfied alongside the Virasoro constraints derived above:
\begin{equation}
\label{eq:nullgaugespectrum}
\begin{aligned}
l_1\left( \lambda - \frac{n_5}{2}w_+\right) + l_2 \left( m_{\rm su}+ \frac{n_5}{2}w_{\rm su}\right) + l_3\frac{E}{2} + l_4 \frac{p_y}{2}=0\,,\\[1mm]
r_1 \left( \bar \lambda + \frac{n_5}{2}w_-\right) + r_2 \left( \bar m_{\rm su} + \frac{n_5}{2}\bar w_{\rm su}\right) + r_3 \frac{E}{2} + r_4 \frac{\bar p_y}{2}=0\,.
\end{aligned}
\end{equation}
Here, we are considering generic states with spectral flow parameters $w_\pm$ on ${SL}(2,\mathbb R)$, $(w_{\rm su},\,\bar w_{\rm su})$ on ${SU}(2)$ and winding $w_y$ on ${ S}^1_y$.

Two relevant subtleties arise when considering spectral flow in the null-gauged WZW model~\cite{Martinec:2018nco,Bufalini:2021ndn}. The first is due to the non-compactness of the time direction $t$: a generic spectral flow along the gauge direction in the holomorphic sector would shift the zero mode of the timelike direction differently with respect to its antiholomorphic counterpart.
Since $t$ is non-compact, this difference is not allowed. This can be enforced by the following condition:
\begin{equation}
l_3 = r_3\,.
\end{equation}
This is consistent with~\eqref{eq:cond1}. Note that holomorphic and antiholomorphic spectral flow on ${SL}(2,\mathbb R)$ are not required to be equal, since asymmetric flows are allowed on the BTZ background~\cite{Hemming:2001we}.

The second subtlety concerns the gauge-triviality of 
spectral flow along the null direction. This implies that a certain combination of spectral flows on ${SL}(2,\mathbb R)$ and ${SU}(2)$, combined with appropriate shifts of the modes $E$, $n_y$ and $w_y$, must leave the physical state invariant. As we show below, this requirement relates the gauging parameters $l_1$ and $r_1$ to the BTZ parameters $\alpha_\pm$, and further imposes quantization conditions on the remaining ones. 

To identify such gauge-equivalent spectral flows, we consider the following shift:
\begin{equation}
w_\pm = \left( \alpha_+\mp \alpha_-\right) \mathbf n \rightarrow  \left( \alpha_+\mp \alpha_-\right) \left(\mathbf n + \mathbf {n_0}\right) \,,\qquad \mathbf{ n_0} \in \mathbb Z\,.
\end{equation}
This shift can be compensated by appropriate changes in the other quantum numbers, leaving invariant the gauge constraints \eqref{eq:nullgaugespectrum} and the Virasoro constraints \eqref{eq:Virasoroholo}, \eqref{eq:Virasoroantiholo}. We therefore take:
\begin{equation}
\left(w_{\rm su},\,\bar w_{\rm su},\, E\,,p_y\,,\bar p_y\right) \rightarrow \left( w_{\rm su} - q_2\,\mathbf{n_0},\,\bar w_{\rm su} - \bar q_2\,\mathbf{n_0},\,E + q_3 \,\mathbf{n_0},\,p_y - q_4 \,\mathbf{n_0},\,\bar p_y - \bar q_4\,\mathbf{n_0}\right)\,.
\end{equation}
To ensure the condition $w_{\rm su} \pm \bar w_{\rm su} \in 2\mathbb Z$ is preserved, we impose $\left(q_2 \pm \bar q_2\right)\mathbf{n_0} \in2 \mathbb Z$ for all $\mathbf{n_0} \in \mathbb Z$. We then introduce two integers $\mathbf{n_{1,2}}\in\mathbb Z$ and require $q_2 = \mathbf{n_1} + \mathbf{n_2}$ and $\bar q_2 = \mathbf{n_1} - \mathbf{n_2}$.\footnote{As explained in~\cite{Bufalini:2021ndn}, when considering fermionic modes this condition becomes more restrictive, requiring $q_2 \in 2\mathbb Z+1$, $\bar q_2 \in 2\mathbb Z +1$.}
Similarly the shifts in $p_y$ and $\bar p_y$ affect the quantized momenta and windings on the circle:
\begin{equation}
n_y \rightarrow n_y - \frac{R_y}{2}\left( q_4 + \bar q_4\right) \mathbf{n_0} \in \mathbb Z\,,\qquad w_y \rightarrow w_y - \frac{1}{2R_y}\left(q_4 - \bar q_4\right) \mathbf{n_0}\in \mathbb Z\,.
\end{equation}
Consequently, we introduce further integers $\mathbf{n_{3,4}}\in\mathbb Z$ and impose: $\left(q_4 - \bar q_4\right)=2 R_y \mathbf{n_3}$ and $ R_y\left( q_4 + \bar q_4\right)=2\mathbf{n_4}$. 

To satisfy the Virasoro constraints \eqref{eq:Virasoroholo}, \eqref{eq:Virasoroantiholo} and the gauge constraints \eqref{eq:nullgaugespectrum} after these shifts, we take the following gauging parameters:
\begin{equation}
\label{eq:quantizationgauging}
\begin{aligned}
l_1 &= \frac{r_3}{q_3}\left(\alpha_+ - \alpha_-\right) \,,\qquad r_1 = -\frac{r_3}{q_3}\left( \alpha_+ + \alpha_-\right)\,,\\[1mm]
l_2 &= \frac{r_3}{q_3}\left( \mathbf{n_1} + \mathbf {n_2}\right)\,,\qquad r_2 = \frac{r_3}{q_3}\left( \mathbf{n_1} - \mathbf {n_2}\right)\,,\\[1mm]
l_4 &= \frac{r_3}{q_3}\left(\frac{\mathbf {n_4}}{R_y}+ \mathbf {n_3} \,R_y \right)\,,\qquad r_4= \frac{r_3}{q_3}\left(\frac{\mathbf {n_4}}{R_y}- \mathbf {n_3} \,R_y \right)\,,
\end{aligned}
\end{equation}
where $q_3$ is given by
\begin{equation}
q_3= \sqrt{\frac{\mathbf {n_4}^2}{R_y^2}+ \mathbf {n_3}^2 \,R_y^2+ n_5\left( \mathbf {n_1}^2 + \mathbf {n_2}^2 + \alpha_+^2 + \alpha_-^2\right)}\,,
\end{equation}
and the integers we introduced satisfy the constraint
\begin{equation}
\mathbf {n_3} \,\mathbf {n_4} + n_5\left( \mathbf {n_1}\,\mathbf {n_2} - \alpha_+ \,\alpha_-\right) =0\,.
\end{equation}
This parametrization is also consistent with the null conditions \eqref{eq:nullconstraint}.\footnote{In the limit $\alpha_+ = 0$, $\alpha_- = -1$ one formally recovers the expressions obtained in~\cite{Bufalini:2021ndn} for the worldsheet model describing the horizonless soliton. In our setup, having independent non-zero values for $\alpha_\pm >0$ allows for more general gauging, so that $l_1 \neq r_1$.}

Using~\eqref{eq:ratio_charges}, the consistency conditions just derived can be translated into constraints on the supergravity charges: 
\begin{equation}
\frac{\mathbf{n_1} - \mathbf {n_2}}{\mathbf {n_1} + \mathbf {n_2}}= \frac{J_+}{J_-}\,,\qquad \qquad \frac{\frac{\mathbf {n_4}}{R_y}- \mathbf {n_3} \,R_y}{\frac{\mathbf {n_4}}{R_y}+ \mathbf {n_3} \,R_y}= \frac{Q_p+ Q_1}{Q_p - Q_1}\,.
\end{equation}
This analysis is consistent with the quantization conditions on the black hole charges, as in~\eqref{eq:charges_params_string}\footnote{Following~\cite{Cvetic:1998xh}, in our conventions the angular momenta are also quantized as $J_1 = -\frac{\mathtt j_\phi}{2}$ and $J_2 = -\frac{\mathtt j_\psi}{2}$, with $\mathtt j_{\phi,\psi}\in \mathbb Z$, consistently with our findings in this section.}
\begin{equation}
    \frac{J_1}{J_2} = -\frac{\mathbf {n_1}}{\mathbf {n_2}}\,,\qquad\qquad \frac{Q_p}{Q_1} = - \frac{{\mathbf{n_4}}}{{\mathbf{n_3} R_y^2}}\,.  
\end{equation}

\medskip

The WZW model describing string propagation on the BTZ background, discussed in appendix~\ref{app:BTZspectrum}, becomes ill-defined in the limit $\alpha_+^2 = \alpha_-^2$, which according to \eqref{eq:quantizationgauging} corresponds to taking an extremal limit ($l_1 r_1 =0$). Therefore, as mentioned above, the model discussed in this section is not reliable for describing extremal black holes, and a separate worldsheet construction is required. Extremal (and supersymmetric) solutions will be the focus of the next section.


\section{Supersymmetric and extremal black holes}
\label{sec:susy}

We are often interested in solutions that preserve some supersymmetries. 
In Lorentzian signature, supersymmetric black holes are necessarily extremal, corresponding in our setup to the limit $\alpha_+^2 \rightarrow \alpha_-^2$. However, as noted earlier, the models constructed so far break down in this regime, where, for instance, the coordinate $\rho$ of \eqref{eq:rhotor} becomes ill-defined. The breakdown of the existing formulation at extremality raises the question, which we address in this section, of how to extend the null-gauging procedure to accommodate the extremal background.

The first step is to introduce an alternative parametrization of the ${SL}(2,\mathbb R)$ group manifold~\cite{Parsons:2009si}:
\begin{equation}
    \label{eq:sl2extremal}
    g_{\rm sl}^{\rm ext} = {\rm e}^{\alpha_0\left(\tau - \sigma\right)\sigma_3}\begin{pmatrix}
        \frac{\sqrt{\alpha_0}}{\sqrt{2}\,\hat r} & -\frac{\hat r}{\sqrt{2\alpha_0}}
        \\
        \frac{\sqrt{\alpha_0}}{\sqrt{2}\,\hat r} & \frac{\hat r}{\sqrt{2\alpha_0}}
    \end{pmatrix}
    {\rm e}^{\frac{\tau + \sigma}{2}\left(\sigma_1 -i\sigma_2\right)}
\end{equation}
Here, the parameter $\alpha_0$ plays the role of the extremal BTZ horizon. This can be seen explicitly by writing the metric on the group manifold in this parametrization, which takes the form
\begin{equation}
\diff s^2 = \kk_{\rm sl} \left[- \frac{\left( r^2 -\alpha_0^2\right)^2}{r^2}\diff \tau^2 + \frac{r^2}{\left( r^2-\alpha_0^2\right)^2}\diff r^2 +  r^2\left( \diff \sigma - \frac{\alpha_0^2}{r^2}\diff \tau \right)^2\right]\,,
\end{equation}
where the new radial coordinate is given by $r^2 = \hat r^2 + \alpha_0^2$. This corresponds to the extremal version ($\alpha_+ = \alpha_-= \alpha_0$) of the BTZ metric in~\eqref{eq:BTZmetric}.

The remainder of this section is structured as follows. In section \ref{sec:susyblackhole}, we analyze the supergravity solution obtained by taking the supersymmetric limit of the black hole presented in section \ref{sec:decoupling}. We demonstrate that imposing supersymmetry alone is not sufficient to ensure a well-defined Lorentzian solution free of closed timelike curves (CTCs) outside and on the horizon. There exist only two well-behaved supersymmetric solutions: one corresponds to the horizonless soliton of~\cite{Giusto:2004id,Giusto:2004ip,Giusto:2004kj,Giusto:2012yz}, which is the supersymmetric version of~\cite{Jejjala:2005yu}\footnote{The null-gauged WZW model to describe string propagation on such a horizonless supersymmetric background was first derived in~\cite{Martinec:2017ztd}, employing an AdS$_3$ factor in global coordinates in the upstairs group manifold.}, valid for $\ell \geq 1$, while the other describes an extremal supersymmetric black hole for $0\leq \ell <1$~\cite{Breckenridge:1996is,Cvetic:1996xz}. The latter is the branch of the solution that we focus on. 

In section \ref{sec:susybackground}, we show how the null gauging procedure can be applied to describe a coset whose target space precisely matches the supersymmetric supergravity background. In the previous section, we showed that the worldsheet currents being gauged can be expressed as a linear combination of current algebra generators (see \eqref{eq:gauging_currents}). For the ${SL}(2,\mathbb R)$ factor, we previously considered the generators ${\cal J}^3_{\rm sl}$ and $\bar {\cal J}^3_{\rm sl}$ acting on the holomorphic and antiholomorphic sectors, respectively. However, here we show that in the extremal case, we must instead act {\it asymmetrically}: the currents specifying the gauging are ${\cal J}_{\rm sl}^{3}$ for the holomorphic sector, and $\bar {\cal J}^-_{\rm sl}= \bar {\cal J}^1_{\rm sl}- \bar {\cal J}^2_{\rm sl}$ for the antiholomorphic one. This asymmetry is the distinctive feature of the extremal case.


\subsection{Supersymmetric black hole in supergravity}
\label{sec:susyblackhole}

In this section, we start from the non-extremal and non-supersymmetric black hole solution discussed in section \ref{sec:decoupling} and impose supersymmetry, following~\cite{Cvetic:1996xz,Cvetic:1997uw}. The solution preserves supersymmetry provided that the mass, given in \eqref{eq:bhcharges}, satisfies a linear relation with the charges:
\begin{equation}
E = Q_1 + Q_5 + Q_p\,,
\end{equation}
which is realized by setting $M = 0$. Under this condition, the metric takes the form
\begin{equation}
\label{eq:susysolution}
\begin{aligned}
{\rm d}s^2 &= \hat\Sigma_0^{-1}\left(-\hat h_{tt}{\rm d}t^2 + \hat h_{yy}{\rm d}y^2\right) - 4Q_p\,\hat\Sigma_0^{-1}\,{\rm d}t \,{\rm d}y \\[2mm]
&\quad + Q_5 \left(\frac{\diff \hat r^2}{\hat r^2 + b\sqrt{\ell^2 -1 }}+ {\rm d}\theta^2\right)+Q_5\,\hat \Sigma_0^{-1}\left[ \hat h_{\phi\phi} \,\sin^2\theta \,{\rm d}\phi^2 +\hat h_{\psi\psi} \,\cos^2\theta \,{\rm d}\psi^2\right] \\[2mm]
&\quad +\hat \Sigma_0^{-1}\left[\sin^2\theta\,\left(\hat \eta_- \, {\rm d}t + \hat \zeta_-\,{\rm d}y\right){\rm d}\phi +\cos^2\theta\,\left(\hat \eta_+ \, {\rm d}t + \hat\zeta_+\,{\rm d}y\right){\rm d}\psi
\right]\,,
\end{aligned}
\end{equation}
where we introduced a new radial coordinate, $\hat r$, related to the one used in \eqref{eq:decoupledmetric} by
\begin{equation}
\sinh^2\rho = \frac{\hat r^2}{r_+^2 - r_-^2}\,,\qquad r_+^2 - r_-^2 = b\sqrt{\ell^2 -1}\,.
\end{equation}
The associated dilaton and B-field are given by
\begin{equation}
\label{eq:susysolution2}
\begin{aligned}
{\rm e}^{2\Phi} &= 2\,Q_5\,\,\hat\Sigma_0^{-1}\,,\\[2mm]
B&= -\frac{2 \,Q_1}{\hat \Sigma_0}\,\diff t\wedge \diff y + \frac{Q_5\,\hat h_{\phi\phi}}{\hat\Sigma_0}\cos^2\theta\, \diff \phi \wedge \diff \psi\\
&\quad +\frac{1}{2\hat\Sigma_0}\left[ \hat \eta_+\,\diff t + \hat\zeta_+\,\diff y\right]\wedge \sin^2\theta\diff\phi + \frac{1}{2\hat\Sigma_0}\left[\hat \eta_- \,\diff t+ \hat \zeta_- \,\diff y\right] \wedge \cos^2\theta \diff\psi\,.
\end{aligned}
\end{equation}
The various functions appearing in the geometry are
\begin{equation}
\begin{aligned}
\hat \Sigma_0&= \left[ 2Q_1+ 2\,\hat r^2 +b\sqrt{\ell^2-1} -  b\,\ell\,\cos2\theta\right]\,,\\[1mm]
-\hat h_{tt}&=\left[2Q_p - 2\,\hat r^2 -b\sqrt{\ell^2-1}+ b\,\ell \,\cos2\theta\right]
\,,\\[1mm]
\hat h_{yy}&=  \left[2Q_p + 2\,\hat r^2 +b\sqrt{\ell^2-1}- b\,\ell \,\cos2\theta\right]
\,,\\[1mm]
\hat h_{\psi\psi}&=  \left[2Q_1+ 2\,\hat r^2 +b\sqrt{\ell^2-1} +  b\,\ell\right]
\,,\\[1mm]
\hat h_{\phi\phi}&= \left[2Q_1+ 2\,\hat r^2 +b\sqrt{\ell^2-1} -  b\,\ell\right]
\,,\\[1mm]
\hat \eta _\pm &= \sqrt{\frac{Q_5}{Q_1\,Q_p}} \Bigl[b(Q_1 + Q_p)\pm 4\ell\, Q_1\, Q_p\Bigr]\,,\\[1mm]
\hat\zeta _\pm &= \sqrt{\frac{Q_5}{Q_1\,Q_p}}\Bigl[b(Q_1 - Q_p) \mp 4\ell\, Q_1 \,Q_p\Bigr]\,.
\end{aligned}
\end{equation}
In the black hole regime, corresponding to $0\leq \ell <1$, some components of the metric become complex, unless we also set $b=0$. This corresponds to the extremal limit, as seen from the vanishing of $r_+^2 - r_-^2$, resulting in the vanishing of one of the two angular momenta appearing in \eqref{eq:bhcharges}.\footnote{Formally, the supersymmetric solution with $b>0$ and $0\leq \ell <1$ describes a supersymmetric black hole with finite temperature. While this solution is pathological in Lorentzian signature, it is possible to make sense of it in Euclidean signature~\cite{Anupam:2023yns}. These supersymmetric, yet ``non-extremal'', Euclidean solutions have been attracting attention recently as they serve as saddles of the gravitational path integral with supersymmetric boundary conditions that computes a supersymmetric index counting microstates. It would be interesting to explore whether the Euclidean version of the corresponding null-gauged model could provide a useful way to study such supersymmetric, non-extremal backgrounds. 
} The well-defined supersymmetric and extremal black hole solution, obtained in the limit $M=b=0$, is the BMPV black hole~\cite{Breckenridge:1996is}. 
In the NS$5$-decoupling limit, using \eqref{eq:bhcharges}, the metric can be expressed as
\begin{equation}
\label{eq:BMPVmetric}
\begin{aligned}
{\rm d}s^2 &= -\frac{\hat r^2 - Q_p}{\hat r^2 + Q_1}\diff t^2 + \frac{\hat r^2 + Q_p}{\hat r^2 + Q_1}\diff y^2 - 2\frac{Q_p}{\hat r^2 + Q_1}\diff t\,\diff y + Q_5 \left( \frac{\diff \hat r^2}{\hat r^2} + \diff \theta^2\right) \\[1mm]
&\quad +Q_5\left( \sin^2\theta \,\diff\phi ^2 + \cos^2\theta\, \diff \psi^2\right) + \frac{J_-}{\hat r^2 + Q_1}\left(\diff t - \diff y\right)\Bigl[ \cos^2\theta\diff \psi-\sin^2\theta \diff\phi \Bigr]
\,,
\end{aligned}
\end{equation}
while the dilaton and B-field read 
\begin{equation}
\begin{aligned}
{\rm e}^{2\Phi} &= \frac{Q_5}{\hat r^2 +Q_1}\,,
\\[1mm]
B&= -\frac{Q_1}{\hat r^2+Q_1}\diff t\wedge \diff y + Q_5\cos^2\theta \diff \phi \wedge \diff \psi +\frac{J_-}{\left( \hat r^2 + Q_1\right)}\frac{ \diff t- \diff y}{2}\wedge \Bigl[ \sin^2\theta \diff \phi - \cos^2\theta\diff \psi\Bigr].
\end{aligned}
\end{equation}
The bound $0\leq \ell <1$ in the supersymmetric regime translates into the inequality
\begin{equation}
Q_1\,Q_5\,Q_p - \frac{J_-^2}{4}>0\,,
\end{equation}
which ensures that the BMPV black hole has a real and positive entropy.

\medskip

On the other hand, in the soliton regime, $\ell >1$,  imposing the supersymmetric version of the no-horizon condition, 
\begin{equation}
b = 4\sqrt{\ell^2 -1}\frac{Q_1\,Q_p}{Q_1 + Q_p}\,,
\end{equation}
ensures the absence of CTCs, leading to the solution found in~\cite{Giusto:2004id,Giusto:2004ip,Giusto:2004kj}. We review the argument confirming the absence of CTCs in appendix \ref{sec:CTC}, by extending to the present setup the analysis of~\cite{Chong:2005hr}.


\subsection{Gauged WZW models at extremality}
\label{sec:susybackground}

In this section, we propose a null-gauged WZW model to describe string propagation on the BMPV black hole background. To achieve this, we start by parametrizing the elements of the upstairs group manifold $\mathcal G^{\rm up}$ \eqref{eq:upgroup} (neglecting the four-torus) as
\begin{equation}
g= \left( g_{\rm sl}^{\rm ext},\, {\rm e}^{\frac{i}{2}\left(- \phi + \psi\right)\sigma_3}{\rm e}^{i\theta \sigma_1}{\rm e}^{\frac{i}{2}\left( \psi + \phi\right)\sigma_3},\,{\rm e}^t,\,{\rm e}^{iy/R_y}\right)\,,
\end{equation}
where $g_{\rm sl}^{\rm ext}$ was given in \eqref{eq:sl2extremal}. The target space metric on the upstairs group manifold takes the form
\begin{equation}
\begin{aligned}
\diff s^2_{\rm up} &= n_5 \left[\frac{\diff \hat r^2}{\hat r^2}  - \left(\hat r^2 - \alpha_0^2\right)\diff \tau ^2 - 2\alpha_0^2\, \diff \tau \diff \sigma + \left(\hat r^2 + \alpha_0^2\right)\diff \sigma^2\right] \\[1mm]
&\quad +n_5\left(\diff \theta^2 + \sin^2\theta \diff\phi^2 + \cos^2\theta \diff\psi^2\right)- \diff t^2 + \diff y^2\,.
\end{aligned}
\end{equation}
This metric corresponds to the upstairs model, characterized by eight holomorphic and eight antiholomorphic conserved worldsheet currents. Following our analysis in section \ref{sec:downstairsmodel}, we want to gauge a certain linear combination of them. To do this explicitly, we first write down the relevant currents. For the ${SL}(2,\mathbb R)$ factor these are given by
\begin{equation}
\begin{aligned}
{\mathcal J}_{\rm sl}^3&= -i \,\kk_{\rm sl}\,{\rm Tr}\left[ t^3_{\rm sl} \, \partial g_{\rm sl}^{\rm ext}\,(g_{\rm sl}^{\rm ext})^{-1}\right] =  -\frac{\kk_{\rm sl}}{2\alpha_0}\left[ \left( \hat r^2 - 2\alpha_0^2 \right) \partial \tau + \left( \hat r^2 + 2\alpha_0^2 \right) \partial\sigma\right]\,
\\[1mm]
\bar {\mathcal J}_{\rm sl}^{-} &= -i \,\kk_{\rm sl}\,{\rm Tr}\left[ \left( t^1_{\rm sl} - t^2_{\rm sl} \right)\,(g_{\rm sl}^{\rm ext})^{-1}\bar \partial g_{\rm sl}^{\rm ext} \right]= \kk_{\rm sl}\, \hat r^2 \left( \bar\partial \tau - \bar\partial \sigma\right)\,,
\end{aligned}
\end{equation}
where the generators of $SL(2,\mathbb R)$ are taken as in \eqref{eq:genSL}. As we anticipated, an asymmetric gauging is required to realize the extremal geometry. For the ${ SU}(2)$ factor, the currents participating in the gauging are
\begin{equation}
\begin{aligned}
\mathcal J_{\rm su}^3 &= -i \,\kk_{\rm su} \,{\rm Tr}\left[ t^3_{\rm su}\,\partial g_{\rm su} \,g_{\rm su}^{-1}\right]= \kk_{\rm su} \left[\cos^2\theta \partial \psi - \sin^2\theta \partial\phi \right]\,,
\\[1mm]
\bar {\mathcal J}_{\rm su}^3 &= -i \,\kk_{\rm su} \,{\rm Tr}\left[ t^3_{\rm su} \, g_{\rm su}^{-1}\,\bar{\partial} g_{\rm su}\right] = \kk_{\rm su} \left[ \cos^2\theta \bar\partial \psi + \sin^2\theta \bar \partial \phi\right] \,.
\end{aligned}
\end{equation}
The final set of currents, corresponding to the directions $t$ and $y$, is simply
\begin{equation}
{\cal P}_t = \partial t\,,\qquad \bar {\cal P}_t = \bar\partial t \,,\qquad {\cal P}_y = \partial y \,,\qquad \bar {\cal P}_y = \bar \partial y\,.
\end{equation}

We now specify the chiral group embeddings that determine the group action being gauged. These are given by
\begin{equation}
\label{extremalembedding}
\begin{aligned}
\varphi_L(h_L) &= \left( -l_1 \sigma_3 \,,\,- i l_2 \,\sigma_3\,,\,l_3 \,,\,- i \frac{l_4}{R_y}\right)h_L\,,\qquad \varphi_R(h_L) = 0\,,\\
\varphi_R(h_R)&= \left( -r_1\left(\sigma_1-i\sigma_2\right)\,,\,ir_2\,\sigma_3\,,\,- r_3 \,,\,i \frac{r_4}{R_y}\right)h_R \,,\qquad \varphi_L(h_R) = 0\,,
\end{aligned}
\end{equation}
where $h_{L,R}$ are some real parameters, and $l_{1,2,3,4}$ and $r_{1,2,3,4}$ are the real gauging coefficients. From these expressions, it is straightforward to verify that the group transformations we wish to gauge correspond to the following action on the coordinates:
\begin{equation}
\begin{aligned}
\delta \tau &= r_1 \,h_R - \frac{l_1}{2\alpha_0}h_L\,,\qquad \delta \sigma = r_1\,h_R + \frac{l_1}{2\alpha_0}h_L\,,\\[1mm]
\delta \phi &= l_2\,h_L-r_2\,h_R \,,\qquad \delta \psi = - \left( r_2\,h_R + l_2 \,h_L\right) \,,\\[1mm]
\delta t& = l_3\,h_L + r_3 \,h_R\,,\qquad \delta y = - \left( l_4 \,h_L + r_4\,h_R\right)\,.
\end{aligned}
\end{equation}
As before, we will use this freedom to fix a convenient gauge in the coset model obtained after implementing the gauging procedure. As discussed in section \ref{sec:downstairsmodel}, the action for the gauged WZW model corresponding to the extremal black hole background is given by \eqref{eq:gaugedsigma} and \eqref{eq:generalgWZWmodel}. The one-forms $\theta_{1,2}$ are now given by
\begin{equation}
\begin{aligned}
\theta_1 &= n_5\left[\frac{l_1}{2\alpha_0}\Bigl(\left( \hat r^2 - 2\alpha_0^2\right) \diff \tau + \left( \hat r^2 + 2\alpha_0^2\right) \diff \sigma \Bigr) + l_2 \left(\sin^2\theta \,\diff \phi - \cos^2\theta \,\diff \psi \right)\right] 
\\[1mm]
&\quad - l_3 \,\diff t - l_4\,\diff y\,,
\end{aligned}
\end{equation}
and
\begin{equation}
\begin{aligned}
\theta_2 =  n_5\left[r_1\,\hat r^2 \left( \diff \tau - \diff \sigma \right) + r_2\left( \cos^2\theta \,\diff\psi + \sin^2\theta \,\diff\phi\right) \right]  + r_3\,\diff t + r_4 \,\diff y\,.
\end{aligned}
\end{equation}
These one-forms correspond to the Killing vectors associated to the actions \eqref{extremalembedding}, which in this case are
\begin{equation}
\label{eq:extgaugingKV}
\begin{aligned}
\xi_1 &= \frac{l_1}{2\alpha_0} \left( -\partial_\tau + \partial_\sigma\right) +l_2 \left( \partial_\phi - \partial_\psi\right) + l_3 \partial_t - l_4 \partial_y\,,\\[2mm]
\xi_2 &= r_1 \left( \partial_\tau + \partial_\sigma\right) - r_2 \left( \partial_\psi +\partial_\phi\right) + r_3 \partial_t - r_4 \partial_y\,.
\end{aligned}
\end{equation}
To ensure these Killing vectors are null, we impose the {\it asymmetric} null constraints
\begin{equation}
\label{eq:extnullconstraint}
\begin{aligned}
n_5\left(l_1^2+l_2^2\right) - l_3^2 + l_4^2=0\,,\qquad \qquad n_5\,r_2^2  - r_3^2 + r_4^2=0\,,
\end{aligned}
\end{equation} 
that are independent of $r_1$. After integrating out the  gauge fields, the action becomes
\begin{equation}
S_{gWZW} = S_{WZW} +\frac{1}{\pi} \int_{\Sigma_2}{\rm d}^2z\,\Sigma^{-1}\,\mathcal J \,\bar{\mathcal J}\,,
\end{equation}
where the function $\Sigma$ is now given by
\begin{equation}
\Sigma = \frac{1}{2}\left[\left( l_3\,r_3 - l_4\,r_4\right) - n_5\left(\frac{l_1\,r_1}{\alpha_0}\hat r^2 + l_2\,r_2\,\cos2\theta \right) \right]\,.
\end{equation}
The gauge currents determined from the one-forms $\theta_{1,2}$ take the form
\begin{equation}
\begin{aligned}
\mathcal J& = l_1\,\mathcal J_{\rm sl}^{3} + l_2 \,\mathcal J_{\rm su}^3 + l_3 \,{\cal P}_t + l_4 \,{\cal P}_y =\\[1mm]
&= -n_5 \left[ \frac{l_1}{2\alpha_0}\Bigl( \left( \hat r^2 - 2\alpha_0^2\right) \partial \tau + \left( \hat r^2 + 2\alpha_0^2\right) \partial\sigma \Bigr) + l_2 \left( \sin^2\theta \, \partial \phi - \cos^2\theta  \partial \psi \right)\right]
\\[1mm]
&\quad + l_3\,\partial t + l_4\,\partial y\,,
\end{aligned}
\end{equation}
for the holomorphic sector, and
\begin{equation}
\begin{aligned}
\bar {\mathcal J} &= r_1\,\bar{\mathcal J}_{\rm sl}^- + r_2 \,\bar {\mathcal J}_{\rm su}^3 + r_3 \,\bar {\cal P}_t + r_4\,\bar {\cal P}_y = \\[1mm]
& = n_5\left[r_1\,\hat r^2 \left( \bar\partial \tau - \bar\partial \sigma \right) + r_2 \left( \cos^2\theta \,\bar\partial \psi + \sin^2 \theta \,\bar\partial \phi \right) \right]+ r_3\,\bar \partial t + r_4 \bar \partial y\,,
\end{aligned}
\end{equation}
for the antiholomorphic part. 


\subsubsection{Supersymmetric black hole from the worldsheet}

After gauge-fixing $\sigma = \tau = 0$, the target space metric of the gauged WZW model can be expressed as
\begin{equation}
\label{eq:extremalbkgr}
\begin{aligned}
{\rm d}s^2 &= \Sigma^{-1}\left(-\hat h_{tt}{\rm d}t^2 + \hat h_{yy}{\rm d}y^2\right) +\left( l_4\,r_3 + l_3\,r_4\right)\Sigma^{-1}\,{\rm d}t \,{\rm d}y \\[1mm]
&\quad+ n_5 \left(\frac{\diff \hat r^2}{\hat r^2}+ {\rm d}\theta^2\right)+n_5\, \Sigma^{-1}\left[ \hat h_{\phi\phi} \,\sin^2\theta \,{\rm d}\phi^2 +\hat h_{\psi\psi} \,\cos^2\theta \,{\rm d}\psi^2\right] \\[1mm]
&\quad+n_5\, \Sigma^{-1}\left[\sin^2\theta\,\left(\hat \eta_- \, {\rm d}t + \hat \zeta_-\,{\rm d}y\right){\rm d}\phi +\cos^2\theta\,\left(\hat \eta_+ \, {\rm d}t + \hat\zeta_+\,{\rm d}y\right){\rm d}\psi
\right]\,,
\end{aligned}
\end{equation}
where 
\begin{equation}
\begin{aligned}
\hat h_{tt} &= -\frac{1}{2}\left[n_5\left(\frac{l_1\,r_1}{\alpha_0}\hat r^2 + l_2\,r_2\cos 2\theta\right) + l_3\,r_3 + l_4\,r_4 \right]\,,\\[1mm]
\hat h_{yy} &=  \frac{1}{2}\left[-n_5\left(\frac{l_1\,r_1}{\alpha_0}\hat r^2 + l_2\,r_2\cos 2\theta\right) + l_3\,r_3 + l_4\,r_4 \right]\,,\\[1mm]
\hat h_{\phi\phi} &=  \frac{1}{2}\left[-n_5\left( \frac{l_1\,r_1}{\alpha_0}\hat r^2 + l_2\,r_2 \right) + l_3\,r_3 - l_4\,r_4 \right] \,,\\[1mm]
\hat h_{\psi\psi} &=  \frac{1}{2}\left[-n_5\left( \frac{l_1\,r_1}{\alpha_0}\hat r^2 - l_2\,r_2 \right) + l_3\,r_3 - l_4\,r_4 \right] \,,\\[1mm]
\hat \eta_\pm & =  r_2 \,l_3\pm r_3\,l_2 \,,\qquad \hat \zeta_\pm =  r_2 \,l_4\pm r_4\,l_2 \,.
\end{aligned}
\end{equation}
The background B-field can be shown to be: 
\begin{equation}
\label{eq:susysolutionB2}
\begin{aligned}
B&= \frac{l_3\,r_4-l_4\,r_3}{2\Sigma}\,\diff t\wedge \diff y - \frac{n_5\,\hat h_{\phi\phi}}{\Sigma}\cos^2\theta\, \diff \phi \wedge \diff \psi\\
&\quad-\frac{n_5}{2\Sigma}\left[ \hat \eta_+\,\diff t + \hat\zeta_+\,\diff y\right]\wedge \sin^2\theta\diff\phi - \frac{n_5}{2\Sigma}\left[\hat \eta_- \,\diff t+ \hat \zeta_- \,\diff y\right] \wedge \cos^2\theta \diff\psi\,.
\end{aligned}
\end{equation}
This background, when the gauging parameters satisfy \eqref{eq:extnullconstraint}, describes an extremal black hole. To confirm this, we analyze the near-horizon geometry following an approach similar to that of section~\ref{sec:causalitytemperature}. First, we rewrite the metric \eqref{eq:extremalbkgr} in an adapted form,
\begin{equation}
 \label{eq:adaptedexttempmetric}
 \begin{aligned}
 \diff s^2 &= - \hat f_{tt}\diff t^2 + n_5 \left( \frac{\diff \hat r^2}{\hat r^2} + \diff \theta^2\right) +n_5\,\Sigma^{-1}\Bigl[n_5^{-1}\,\hat h_{yy}\left( \diff y -\hat \omega_y \diff t\right)^2 
 \\[1mm]
 & \quad +\hat h_{\phi\phi} \sin^2\theta\left( \diff \phi - \hat \omega_\phi \diff t\right)^2 +\hat h_{\psi\psi} \cos^2\theta \left(\diff\psi - \hat \omega_\psi \diff t\right)^2
 \\[1.5mm]
 &\quad +\hat \zeta_-\,\sin^2\theta\left( \diff \phi - \hat \omega_\phi \diff t\right)\left( \diff y - \hat \omega_y \diff t\right)+\hat \zeta_+\cos^2\theta \left(\diff\psi - \hat \omega_\psi \diff t\right)\left( \diff y -\hat \omega_y \diff t\right)\Bigr]\,.
 \end{aligned}
 \end{equation}
The undetermined functions appearing in this ansatz can be directly extracted by comparing \eqref{eq:adaptedexttempmetric} with \eqref{eq:extremalbkgr}. Near $\hat r \sim 0$, where the $\diff \hat r^2$ component of the line element exhibits a double pole, we expand the metric as
 \begin{equation}
 \diff s^2 = \frac{n_5}{4}\left( - \hat \rho^2 \,\diff t^2 + \frac{\diff \hat \rho^2}{\hat \rho^2}\right) + ... \,,\qquad \hat r^2 = \frac{\alpha_0\sqrt{r_4^2+ n_5\,r_2^2}}{2r_1}\hat \rho\,.
\end{equation}
This reveals an AdS$_2$ factor in the near-horizon geometry, confirming that the background describes an extremal black hole.

\medskip

To establish the correspondence between the null-gauged WZW model and the BMPV black hole, we now compare our target space geometry \eqref{eq:extremalbkgr} to the extremal and supersymmetric supergravity solution \eqref{eq:BMPVmetric}. To this end, we choose the gauging parameters as follows (recalling that $n_5 = Q_5$):
\begin{equation}
\begin{aligned}
l_1 &\,=\, -\frac{2\alpha_0}{r_1}\frac{l_3^2}{Q_5\left(Q_1+Q_p\right)}\,,\qquad l_2 \,=\, l_3\frac{J_-}{Q_5\left(Q_1 + Q_p\right)}\,, \qquad l_4 \,=\, l_3\frac{Q_1-Q_p}{Q_1+Q_p}\,,
\\[2mm]
r_2 &\, =\, 0\,, \qquad\qquad r_3 \,=\, l_3\,=\, \frac{r_1}{2\alpha_0}\sqrt{4 Q_1 Q_p Q_5 - J_-^2}\,, \qquad \qquad   r_4 \,=\,-l_3\,.
\end{aligned}
\end{equation}
This parametrization has been chosen to satisfy the asymmetric null constraints \eqref{eq:extnullconstraint}. Also, note that the gauging coefficients remain real as long as $4 Q_1\,Q_5\,Q_p > J_-^2$, which holds in the black hole regime (as ensured by the absence of CTCs outside and on the horizon, see app.~\ref{sec:CTC}). Unlike the cases studied in sections~\ref{sec:worldsheet} and~\ref{sec:stringspectrum}, not all gauging parameters are here determined in terms of the independent supergravity variables, since $r_1$ remains arbitrary (though it must be non-zero).  

It would be interesting to perform a detailed analysis of the spectrum of the null-gauged WZW model constructed in this section. 
One would need to study the general Virasoro constraints and the asymmetric gauge constraints that relate quantum numbers and spectral flow parameters to the gauging data, in analogy with section~\ref{sec:stringspectrum}. While Type II superstring theory on BTZ$\times S^3 \times T^4$ in a hyperbolic parametrization has been considered for instance in~\cite{Rangamani:2007fz}, to our knowledge, no complete study has been carried out in the asymmetric hyperbolic/parabolic parametrization of~\eqref{eq:sl2extremal} (see~\cite{Parsons:2009si} for a discussion), nor in the presence of an asymmetric gauging of the type considered here. We plan to explore these directions in the near future.


\section{Discussion}
\label{sec:discussion}

String theory on certain asymptotically linear dilaton backgrounds admits a worldsheet description in terms of null-gauged WZW models. In this paper we analysed the three-charge NS$5$-F$1$-P configuration in a regime where the supergravity solution possesses a Killing horizon, in the NS$5$-decoupling limit which decouples the asymptotically flat region. The construction of such worldsheet models starts from a WZW model on a twelve-dimensional group manifold ${\cal G}^{\rm up}= SL(2,\mathbb R)\times SU(2) \times \mathbb R\times U(1)$; we then gauge two null linear combinations of the Cartan currents of $SL(2,\mathbb R)\times SU(2)$ and the momenta along $\mathbb R\times U(1)$. We highlight the role of different parametrizations of $SL(2,\mathbb R)$ in this context: we employ a parametrization of $SL(2,\mathbb R)$ adapted to a rotating non-extremal BTZ factor, corresponding to a hyperbolic basis for its current algebra.
The gauging is characterised by eight parameters specifying the embedding of the $U(1)_L\times U(1)_R$ group into ${\cal G}^{\rm up}$. These must obey two null constraints, ensuring anomaly cancellation. Moreover, by analysing the Virasoro constraints and additional gauge constraints that select physical states in the gauged model, we derive four quantization conditions on the gauging parameters. The two remaining gauging parameters are then related to the mass and angular momentum of the BTZ factor (equivalently, to the positions of its inner and outer horizons). 

We observed that the worldsheet model describing the JMaRT horizonless NS$5$-F$1$-P configuration constructed in previous literature (see e.g.~\cite{Martinec:2017ztd,Martinec:2018nco,Bufalini:2021ndn}) is recovered in the limit where the inverse Hawking temperature of the background is sent to zero, while also taking certain analytic continuations of the gauging parameters.

Finally, we examined a supersymmetric, extremal limit in which the three-charge NS$5$-F$1$-P configuration we consider reduces to the ten-dimensional uplift of the BMPV black hole (in the NS$5$-decoupling limit). We proposed a corresponding null-gauged WZW model, constructed in analogy to the non-extremal case but adapted to the mixed hyperbolic/parabolic parametrization of $SL(2,\mathbb R)$ that leads to an extremal, rotating BTZ factor.

The class of null-gauged WZW models considered here provides concrete examples of solvable worldsheet CFTs and offers a natural framework to probe string dynamics beyond the black hole near-horizon limit.
By adapting the analysis of~\cite{Martinec:2018nco,Martinec:2019wzw,Martinec:2020gkv,Bufalini:2021ndn,Dei:2024uyx,Martinec:2025xoy}, it should be possible to determine the full physical spectrum of these models. As briefly summarized above, Sec.~\ref{sec:stringspectrum} takes an initial step in this direction for the models associated with non-extremal backgrounds. However, a systematic classification of the physical states, together with an analogous treatment of the models introduced in Sec.~\ref{sec:susy}, is left for future work. 
It would be particularly interesting to investigate dynamical processes such as absorption, reflection, scattering, and the emission of string states, extending the analysis of~\cite{Dijkgraaf:1991ba} for the two-dimensional black hole~\cite{PhysRevD.44.314} (see~\cite{Martinec:2023plo} for a recent study of winding mode emission in the BTZ background). 

\medskip

Another interesting direction is to investigate the relation between the null-gauging of WZW models involving an $SL(2,\mathbb R)$ factor and solvable irrelevant $T\bar T$ deformations of these backgrounds. These deformations provide interesting examples of non-AdS holography. In particular, it was shown in~\cite{Giveon:2017myj} that the NS$5$-F$1$ background of~\cite{Giveon:1999zm}, dual to certain single-trace $T\bar T$-deformed CFTs (see~\cite{Giveon:2017nie} for a discussion), can be formulated as a null-gauged WZW model. To show the relation with our formulation we briefly sketch this construction.
The upstairs model is defined on $SL(2,\mathbb R)\times \mathbb R\times U(1)$ (neglecting the three-sphere, which plays no role here), with $SL(2,\mathbb R)$ parametrized in Poincaré coordinates as
\begin{equation}
    g_{\rm sl} = {\rm e^{\frac{\sigma-\tau}{2}\left( \sigma_1 - i \sigma_2\right)}{\rm e}^{\rho \sigma_3}{\rm e}^{\frac{\sigma+ \tau}{2}\left( \sigma_1 + i \sigma_2\right)}}
\end{equation}
giving the metric of the massless extremal BTZ black hole
\begin{equation}
    \diff s_{SL(2,\mathbb R)}^2 = \mathtt k_{\rm sl} \Bigl[ \diff\rho^2 + {\rm e}^{2\rho}\left( -\diff \tau^2 + \diff \sigma^2\right)\Bigr]\,.
\end{equation}
We then gauge the null currents 
\begin{equation}
\begin{aligned}
    {\cal J} &= l_1 {\cal J}_{\rm sl}^- + l_3\left( \partial t + \partial y\right)\,,\qquad\qquad {\cal J}_{\rm sl}^-= -i \,{\rm Tr}\Bigl[\left(t^1_{\rm sl} - t^2_{\rm sl}\right)\partial g_{\rm sl}\,g_{\rm sl}^{-1} \Bigr]\,,
    \\[1mm]
    \bar {\cal J} &= l_1 \bar {\cal J}_{\rm sl}^+ + l_3\left(\partial t - \partial y\right)\,,\qquad \qquad \bar {\cal J}_{\rm sl}^+ = -i\,{\rm Tr}\Bigl[\left( t^1_{\rm sl} + t^2_{\rm sl}\right) g_{\rm sl}^{-1}\,\bar \partial g_{\rm sl} \Bigr]\,.
    \end{aligned}
\end{equation}
The resulting target space geometry is given by
\begin{equation}
    \diff s^2 = n_5 \left[\lambda^2\, \frac{-\diff t^2 + \diff y^2}{{\rm e}^{-2\rho} + n_5\lambda^2} + \diff \rho^2 + \diff \theta^2 + \sin^2\theta \diff\phi^2 + \cos^2\theta \diff \psi^2\right]\,,\qquad \lambda = \frac{l_1}{l_3}\,,
\end{equation}
corresponding to the two-charge ($Q_p=0$) limit of the BMPV black hole \eqref{eq:BMPVmetric} by taking $\lambda^{-2} = Q_1Q_5$. As we discussed in section~\ref{sec:sugra} this background interpolates between an AdS$_3$ and a linear dilaton region. Intuitively, in the first regime ($\lambda \ll 1$) one effectively gauges away the momenta along $\mathbb R\times U(1)$, recovering AdS$_3$ whose spectrum is that of the dual undeformed CFT$_2$. In the opposite regime, the $SL(2,\mathbb R)$ currents are gauged away, yielding a linear dilaton background associated with a Hagedorn growth, analogous to a $T\bar T$-deformed CFT$_2$. As mentioned before, a similar relation between the supergravity background we discussed in section~\ref{sec:mald_example} and $T\bar T$-deformed CFT$_2$ was observed in\cite{Chakraborty:2020swe,Apolo:2019zai}. In this work, we provided a worldsheet description of that background as a null-gauged WZW model, obtained from the general class in section~\ref{sec:worldsheet} by turning off the rotation parameters ($b= \ell = 0$). One would like to further elucidate the connection between these two descriptions, and additionally to investigate whether more general deformations, such as those in~\cite{Chakraborty:2019mdf,Apolo:2021wcn}, can also be realized as null-gauged WZW models. In this context, the analysis of correlators in null-gauged WZW models (see for instance \cite{Bufalini:2022wzu,Bufalini:2022wyp}) could be useful to further investigate these examples of non-AdS holography.

\medskip

A natural next step would be to develop the Euclidean version of the null-gauged WZW models presented here.
The Euclidean formulation is well-suited for computing the one-loop partition function through a Euclidean path integral, from which the spectrum of the theory can be extracted~\cite{Gawedzki:1991yu,Maldacena:2000kv}. For the coset CFTs of interest, the partition function can likely be obtained by extending the results of~\cite{Dei:2024uyx}. It was recently shown in~\cite{Ferko:2024uxi} that the one-loop string partition function on Euclidean (non-supersymmetric) BTZ at low temperatures reproduces the contribution of the Schwarzian mode, which is responsible for large quantum corrections for near-extremal black holes. It would therefore be interesting to compute the one-loop partition function of the null-gauged WZW models constructed here and analyze the low-temperature limit, thus extending the results of~\cite{Ferko:2024uxi} to our setup.
Also, when the coset CFT is given by a null-gauged supersymmetric WZW model, it may be possible to determine the partition function by applying localization techniques, building on the recent developments of~\cite{Murthy:2025ioh}.  

The Euclidean spectrum exhibits stringy features that are invisible in the supergravity description. In particular, in Euclidean black holes strings can wind around the compact directions, including the Euclidean time. 
When a closed string winds around a circle whose size is shorter than the string length, tachyonic winding modes may appear. Winding tachyons have been considered in both Lorentzian and Euclidean BTZ geometries (see e.g.~\cite{Rangamani:2007fz,Berkooz:2007fe}), where their condensation has been linked to phenomena such as phase transitions and topology change (cf.~\cite{Berkooz:2007fe}).  Additionally, Euclidean black holes develop a cigar geometry, obtained by foliating the orbits of the thermal circle over the radial direction. Closed strings can wind around this cigar, and at its tip (corresponding to the would-be horizon) winding modes can condense. Such winding condensates have been conjectured to account for (at least part of) the black hole entropy (cf.~\cite{Chen:2021dsw,Halder:2023adw,Brustein:2021cza,Brustein:2022wiq} and references therein). A relevant question is whether similar winding-sector effects arise in the null-gauged WZW models constructed here, and whether they offer a useful way to interpret the $\beta \to 0$ limit discussed above.  
In~\cite{Jafferis:2021ywg}, an explicit stringy realization of the ER=EPR correspondence for $AdS_3$ black holes was proposed by considering a Lorentzian continuation of the FZZ duality. It might be possible to extend this proposal to the more general black hole solutions considered in this paper.

Finally, one would like to explore whether our formulation of null-gauged WZW models can be generalized to other string backgrounds. For instance, an interesting case is superstring theory on AdS$_3 \times S^3 \times S^3 \times S^1$\cite{Elitzur:1998mm}, which has recently attracted renewed attention\cite{Murthy:2025moj,Heydeman:2025fde}, or on AdS$_3\times (S^3/Z_{\mathbb N}) \times T^4$~\cite{Emparan:2025ymx}.
We plan to return to these questions in future work.


\section*{Acknowledgments}

We thank Luca Novelli for collaboration on related topics. We thank Maxim Emelin, Davide Cassani, Emil Martinec, Sameer Murthy and David Turton for interesting discussions. We are grateful to Alejandro Ruipérez for comments on the draft.


\appendix


\section{Review of gauged WZW models}
\label{app:WZWmodels}

We briefly review here some general aspects of the null-gauging formalism for sigma models, with emphasis on its application to WZW models. References for this part include~\cite{Hull:1989jk,Tseytlin:1993my,Klimcik:1994wp,Figueroa-OFarrill:2005vws} (see also~\cite{Martinec:2019wzw,Martinec:2020gkv,Bufalini:2021ndn}). We use units in which $\alpha' = 1$. 

\subsection{Gauged non-linear sigma-models}

Consider a $D$-dimensional spacetime $\mathcal M$ with coordinates $X^M$, $M=0,1,\dots,D-1$. To describe string propagation on this background, we regard $X^M$ as maps $X^M:\Sigma_2 \to \mathcal M$, where $\Sigma_2$ is the string worldsheet. Their dynamics is governed by a worldsheet sigma model with action
\begin{equation}
S_\sigma =\frac{1}{2\pi} \int_{\Sigma_2}\,g_{MN}(X)\, \diff X^M \wedge \star_2\,\diff X^N\,,
\end{equation}
where $g_{MN}$ is the spacetime metric on the target space ${\cal M}$, and $\star_2$ denotes the Hodge-star operator on $\Sigma_2$.

If the string couples electrically to a $B_{(2)}$-field, a Wess–Zumino (WZ) term can be included. If $\Sigma_2$ is compact, this can be written as an integral over a three-dimensional manifold $\Omega_3$, such that $\partial \Omega_3 = \Sigma_2$, in terms of a globally-defined, closed three-form $H_{(3)}$: 
\begin{equation}
S_{WZ} = \frac{1}{\pi}\int_{\Omega_3}H_{(3)} = \frac{1}{6\pi}\int_{\Omega_3}H_{MNP}(X) \,{\rm d} X^M\wedge {\rm d}X^N \wedge {\rm d}X^P\,.
\end{equation}
If $H_{(3)}$ is also exact, $H_{(3)} = \diff B_{(2)}$ for a globally-defined $B_{(2)}$, the WZ term reduces to a worldsheet integral,
\begin{equation}
S_{WZ} = \frac{1}{4\pi}\int_{\Sigma_2}B_{MN}(X) \,{\rm d} X^M\wedge {\rm d}X^N\,.
\end{equation}
Isometries of the target space ${\cal M}$, generated by the Killing vectors $\xi^M_a$, correspond to global symmetries of the sigma model, acting on $X^M$ as $\delta X^M = \lambda_a \xi^M_a$ with constant parameters $\lambda_a$. These symmetries can be gauged by promoting the parameters to local functions on $\Sigma_2$ and coupling to corresponding gauge fields, i.e. worldsheet one-forms valued in the Lie algebra $\mathfrak g$ of the isometry group $\mathcal G$. For each generator $u_a \in \mathfrak g$, there is an associated vector field $\xi^M_a$ on $\mathcal M$ satisfying
\begin{equation}
    {\cal L}_{\xi_a}H_{(3)}=0\,,
\end{equation}
where $\mathcal L_{\xi_a} = \diff \iota_{\xi_a} + \iota_{\xi_a}\diff$ is the Lie derivative. This holds since $H_{(3)}$ is closed and invariant. This condition also implies the existence of a set of one-forms $\theta_a$ on $\mathcal M$ such that~\cite{Hull:1989jk}
\begin{equation}
\label{eq:gaugingconsistency1}
\iota_{\xi_a}H_{(3)}= \diff \theta_a\,.
\end{equation}
We want to gauge a subgroup $\mathcal H \subset \mathcal G$. To do so, we couple the sigma model to a set of worldsheet gauge fields $\mathcal A^{\hat a}$, transforming under the local symmetry, with $\hat a = 1,\,...\,.{\rm dim}{\cal H}$. The gauged sigma model term is obtained by substituting ordinary derivatives with covariant ones
\begin{equation}
S_\sigma \to \frac{1}{2\pi}\int_{\Sigma_2}g_{MN} \,DX^M \wedge \star_2DX^N\,,
\end{equation}
with $DX^M = \diff X^M - \xi^M_{\hat a}{\cal A}^{\hat a}$. On the other hand, a consistent gauging of the WZ term is achieved only if $H_{(3)}$ extends to an equivariant closed form~\cite{Figueroa-OFarrill:2005vws}
\begin{equation}
    H_{(3)} \to H_{(3)} + \diff \left({\cal A}^{\hat a}\wedge \theta_{\hat a} + \frac{1}{2}\iota_{\xi_{\hat a}}\theta_{\hat b} \,{\cal A}^{\hat a}\wedge {\cal A}^{\hat b}\right)\,,
\end{equation}
with one-forms $\theta_{\hat a}$ satisfying 
\begin{equation}
\label{eq:gaugingconsistency2}
    \iota_{\xi_{\hat a}}\theta_{\hat b} + \iota_{\xi_{\hat b}}\theta_{\hat a} = 0\,.
\end{equation}
Putting both terms together, a consistent gauged model reads
\begin{equation}
 \label{eq:gaugedsigma}
\begin{aligned}
S_{gWZ} &=  \frac{1}{2\pi}\int_{\Sigma_2}\Bigl[ g_{MN}\, D X^M \wedge \star_2 DX^N + {\cal A}^{\hat a}\wedge \theta_{\hat a} +\frac{1}{2} \iota_{\xi_{\hat a}}\theta_{\hat b} \,{\cal A}^{\hat a}\wedge {\cal A}^{\hat b}\Bigr]
\\[1mm]
&\quad +\frac{1}{\pi}\int_{\Omega_3}H_{(3)}\,.
\end{aligned}
\end{equation} 


\subsection{Null-gauged WZW models}

We will now consider the case in which the target space for the ungauged model, denoted by ${\cal M}^{\rm up}$, is the Lie group ${\cal G}^{\rm up}$. The total isometry group of $\mathcal M^{\rm up}$ is given by $\mathcal G^{\rm up}_L \times \mathcal G^{\rm up}_R$, acting on $\mathcal G^{\rm up}$ as
\begin{equation}
g \rightarrow g_L \,g\, g_R^{-1}\,,\qquad g \in {\cal G}^{\rm up}\,,\qquad g_{L,R} \in \mathcal G^{\rm up}_{L,R} \,.
\end{equation}
Left and right Maurer-Cartan one-forms encode the structure of the group manifold, and they are respectively given by\begin{equation}
\theta^L = g^{-1}\,{\rm d}g\,, \qquad \qquad \theta^R= -{\rm d}g\,g^{-1}\,.
\end{equation}
The standard bi-invariant metric on $\mathcal M^{\rm up}$ is, then, given by
\begin{equation}
{\rm d}s^2 =g_{MN}\,{\rm d}X^M\,{\rm d}X^N \to {\rm sgn}\,\frac{\mathtt k}{2}\,\text{Tr}\left[ \left( g^{-1}\,{\rm d}g\right)^2\right]\,,
\end{equation}
and the bi-invariant three form is
\begin{equation}
H_{(3)} = {\rm sgn}\,\frac{\mathtt k}{3!}\,\text{Tr} \left[ \theta^L \wedge \theta^L \wedge \theta^L\right]\,,
\end{equation}
where $\mathtt k \in \mathbb Z$ is the level of the associated current algebra (we will take ${\rm sgn}=1$ for $SL(2,\mathbb R)$, while ${\rm sgn} =-1$ for $SU(2)$, $U(1)$ and $\mathbb R$). 

Therefore, the WZW model on ${\cal M}^{\rm up}$ is given by
\begin{equation}\label{eq:upstairsmodel}
S_\sigma + S_{WZ} = {\rm sgn}\frac{\mathtt k}{2\pi}\left( \int_{\Sigma_2}{\rm d}^2z\,\text{Tr}\left[ \partial g\, g^{-1} \overline \partial g\,g^{-1}\right]+ \frac{1}{3}\int_{\Omega_3}\text{Tr}\left[ \left( g^{-1}\,{\rm d}g\right)^3\right] \right)\,.
\end{equation}
In order to discuss the gauging of the subset $\mathcal H \subset \mathcal G^{\rm up}$, we need to specify how this group is embedded into the isometry group $\mathcal G^{\rm up}_L \times \mathcal G^{\rm up}_R$. The embedding is specified in terms of a pair of homomorphisms $\varphi_{L,R} : \mathcal H \rightarrow \mathcal G^{\rm up}_{L,R}$. In other words, the action to be gauged is given by
\begin{equation}
g \rightarrow \varphi_L(h_0)\, g\, \varphi_R(h_0)^{-1}\,,\qquad h_0 \in \mathcal H\,.
\end{equation}  
The group embeddings ${\varphi}_{L,R}$ induce corresponding Lie algebra homomorphisms, which we also denote by $\varphi_{L,R}$ by an abuse of notation. Let $h_{\hat a}$ be a basis for the Lie algebra of ${\cal H}$, denoted by $\mathtt h$. 
Then, for each $h_{\hat a}$ there is a corresponding Killing vector
\begin{equation}
    \xi_{\hat a} = - \left( \varphi_L\left( h_{\hat a}\right)\right)^R - \left( \varphi_R\left(h_{\hat a}\right)\right)^L\,.
\end{equation}
Vectors of the form $X^{L,R}$ denote the left/right-invariant vector fields. The action of the Maurer-Cartan forms on them is given by $\theta^{L,R}\cdot X^{L,R} = X$. With these ingredients, one can show that \eqref{eq:gaugingconsistency1} can be solved by~\cite{Figueroa-OFarrill:2005vws} 
\begin{equation}
\begin{aligned}
\theta_{\hat a} &= \langle \varphi_R(h_{\hat a}),\, \theta^L \rangle- \langle \varphi_L(h_{\hat a}),\,\theta^R\rangle
\\[1mm]
&= \langle \varphi_R(h_{\hat a}),\,g^{-1}{\rm d}g\rangle + \langle \varphi_L (h_{\hat a}),\,{\rm d}g\,g^{-1}\rangle\,,
\end{aligned}
\end{equation}
where we introduced the notation of inner product for matrix groups, whose normalization is given by $\langle A,\,B\rangle \equiv {\rm sgn}\,\frac{\mathtt k}{2}{\rm Tr}\left[ A \,B\right]$. 

Then, the consistency condition \eqref{eq:gaugingconsistency2} is satisfied provided
\begin{equation}
\text{Tr}\Bigl[ \varphi_L (h_{\hat a})\varphi_L (h_{\hat b})\Bigr] -\text{Tr}\Bigl[ \varphi_R(h_{\hat a})\varphi_R (h_{\hat b})\Bigr]=0\,.
\end{equation} 
We will in particular consider the case in which the group to be gauged is given (at least locally) by $\mathcal H = U(1)_L \times U(1)_R$. Then, a basis for the generators of the Lie algebra associated to $\mathcal H$ is simply given by a pair of real numbers $h_{1,2}\in \mathbb R$. Consistent embeddings are obtained by taking
\begin{equation}
\label{eq:chiral_embedding}
\varphi_R (h_1) = 0\,,\qquad \varphi_L(h_2) = 0\,, \qquad {\rm Tr}\Bigl[ \varphi_R(h_2)^2\Bigr]= {\rm Tr}\Bigl[ \varphi_L(h_1)^2\Bigr]=0\,.
\end{equation}
These translate into the following choice for the one-forms $\theta_{1,2}$:
\begin{equation}
\begin{aligned}
\theta_1 &= {\rm sgn}\,\frac{\mathtt k}{2}  {\rm Tr}\left[ \varphi_L(h_1)\, {\rm d}g \,g^{-1}\right]\,,
\\[1mm]
\theta_2& = {\rm sgn}\,\frac{\mathtt k}{2} {\rm Tr}\left[ \varphi_R(h_2)\, g^{-1}\,{\rm d}g\right] \,,
\end{aligned}
\end{equation}
whose components can be related to the vectors specifying the gauging,
\begin{equation}
\theta_1{}_M = g_{MN}\,\xi_1^N\,, \qquad \theta_2{}_M = -g_{MN}\,\xi_2^N\,.
\end{equation}
Additionally, the above conditions require the Killing vectors to be null
\begin{equation}
\xi^M_{\hat a} \xi_M{}_{\hat a} = 0\,,\qquad \hat a=1,2\,.
\end{equation}
Going back to \eqref{eq:gaugedsigma}, introducing a new notation for the independent components of the gauge fields,
\begin{equation}
    {\cal A}^1_z = {\cal A}'\,,\qquad {\cal A}^2_z= {\cal A}\,,\qquad {\cal A}^1_{\bar z} = \bar {\cal A}\,,\qquad {\cal A}^2_{\bar z} = \bar {\cal A}'\,,
\end{equation}
one can notice that, since the gauging \eqref{eq:chiral_embedding} is chiral and null, half of these components decouple, and the resulting WZW action just depends on ${\cal A}$ and $\bar {\cal A}$~\cite{Martinec:2019wzw}, 
\begin{equation}
\label{eq:generalgWZWmodel}
\begin{aligned}
S_{gWZW} &= {\rm sgn}\,\frac{\mathtt k}{2\pi}\Bigl( \int_{\Sigma_2}{\rm d}^2z\,\text{Tr}\left[ \partial g\, g^{-1} \overline \partial g\,g^{-1}\right]+ \frac{1}{3}\int_{\Omega_3}\text{Tr}\left[ \left( g^{-1}\,{\rm d}g\right)^3\right] \Bigr)
\\[1mm]
&\quad +\frac{1}{\pi} \int_{\Sigma_2}{\rm d}^2z\left[ 2 \mathcal A\, \theta_2{}_M\, \overline {\partial}X^M - 2 \bar{\mathcal A}\,\theta_1{}_M \, \partial X^M - 4\Sigma\, \mathcal A\, \bar {\mathcal A}\right]\,,
 \end{aligned}
\end{equation}
where
\begin{equation}
\Sigma= -\frac{1}{2}\xi^M_1 \,g_{MN}\, \xi^N_2\,.
\end{equation}
The gauged WZW action \eqref{eq:generalgWZWmodel} provides a sigma model on the coset 
\begin{equation}
\frac{\mathcal G^{\rm up}}{U(1)_L \times U(1)_R}\,.
\end{equation}
Since the background gauge fields enter the action  quadratically, they can be easily integrated out. 
While this provides a target space description that may receive $1/\mathtt{k}$ corrections, it remains directly comparable to the corresponding two-derivative supergravity solution, which is also valid at leading order in the semiclassical approximation.

In order to express the final result in a convenient way, let us introduce the worldsheet currents
\begin{equation}
\begin{aligned}
\mathcal J =- \theta_1{}_M\,\partial X^M= -{\rm sgn}\,\frac{\mathtt k}{2}  {\rm Tr}\left[ \varphi_L(h_1)\, \partial g \,g^{-1}\right] \,,\\[1mm]
\bar {\mathcal J} = \theta_2{}_M\, \overline {\partial} X^M= {\rm sgn}\,\frac{\mathtt k}{2} {\rm Tr}\left[ \varphi_R(h_2)\, g^{-1}\,\overline\partial g\right]\,.
\end{aligned}
\end{equation}
After integrating out the gauge fields, the terms in the action \eqref{eq:generalgWZWmodel} simply reduce to
\begin{equation}\label{eq:gaugedaction}
\frac{1}{\pi}\int_{\Sigma_2}{\rm d}^2z\,\Sigma^{-1}\mathcal J\, \bar{\mathcal J}\,.
\end{equation}
The overall effect of the gauging therefore is to add a term \eqref{eq:gaugedaction} to the ungauged action \eqref{eq:upstairsmodel}, resulting in the model
\begin{equation}
\label{eq:gaugedWZWaction}
\begin{aligned}
S_{gWZW}& ={\rm sgn}\frac{\mathtt k}{2\pi}\Big( \int_{\Sigma_2}{\rm d}^2z\,\text{Tr}\left[ \partial g\, g^{-1} \overline \partial g\,g^{-1}\right]+ \frac{1}{3}\int_{\Omega_3}\text{Tr}\left[ \left( g^{-1}\,{\rm d}g\right)^3\right]\Big)\\[1mm]
&\quad +\frac{1}{\pi} \int_{\Sigma_2}{\rm d}^2z\,\Sigma^{-1}\mathcal J\, \bar{\mathcal J}\,.
\end{aligned}
\end{equation}


\section{Strings on BTZ background}
\label{app:BTZspectrum}

In this appendix, we review aspects of string propagation on the background BTZ $\times$ $\mathcal M_{\rm int}$, where $\mathcal M_{\rm int}$ is a generic internal CFT. Our presentation closely follows the analyses in~\cite{Natsuume:1996ij,Hemming:2001we}, emphasizing features relevant to our main discussion. The BTZ black hole arises as a discrete orbifold of AdS$_3$, which can be described via a specific parametrization of the ${SL}(2,\mathbb{R})$ group element:
\begin{equation}
\label{eq:BTZgroupmanifold}
g_{\rm sl} = {\rm e}^{\frac{1}{2}\left( \tau + \sigma\right) \sigma_3} {\rm e}^{\rho \sigma_1} {\rm e}^{\frac{1}{2}\left( \sigma - \tau \right) \sigma_3}\,.
\end{equation}
This coordinate patch, valid for $\rho \geq 0$, describes the spacetime region outside the outer horizon of the BTZ black hole. The target space metric derived from the WZW model reads:
\begin{equation}
\diff s^2 = \kk \left[ - \sinh^2\rho \,\diff \tau^2 + \diff\rho^2+ \cosh^2\rho\, \diff \sigma^2\right]\,.
\end{equation}
The orbifold action that gives a rotating BTZ is realized via the coordinate identifications
\begin{equation}
\label{eq:btzglobalidentifications}
\left( \tau ,\,\sigma\right) \sim \left( \tau - 2\pi \alpha_- ,\,\sigma + 2\pi \alpha_+\right)\,,
\end{equation}
with real parameters $\alpha_\pm$ satisfying $\alpha_+ > \alpha_-$. To reveal the BTZ structure more clearly, we introduce a new coordinate system:
\begin{equation}
\label{eq:rhotor}
\sinh^2 \rho = \frac{r^2 - \alpha_+^2}{\alpha_+^2 - \alpha_-^2}\,,\qquad \tau = \alpha_+ \,t - \alpha_- \, \phi \,,\qquad \sigma= -\alpha_- \,t + 
\alpha_+ \,\phi\,,
\end{equation}
under which the metric takes the standard form:
\begin{equation}
\label{eq:BTZmetric}
\diff s^2 = \kk \left[- \frac{\left( r^2 - \alpha_+^2\right)\left( r^2 - \alpha_-^2\right)}{r^2}\diff t^2 + \frac{r^2}{\left( r^2 - \alpha_+^2\right)\left( r^2 - \alpha_-^2\right)}\diff r^2 + r^2\left( \diff \phi - \frac{\alpha_-\,\alpha_+}{r^2}\diff t\right)^2 \right]\,.
\end{equation}
This corresponds to the BTZ black hole metric with mass and angular momentum:
\begin{equation}
M_{\rm BTZ} = \alpha_+^2 + \alpha_-^2 \,,\qquad J_{\rm BTZ} = 2 \alpha_- \,\alpha_+\,.
\end{equation}
The identification $\phi \sim \phi + 2\pi$ defines the quotient structure, while $t \in \mathbb{R}$ indicates that we are working on the universal cover of ${SL}(2,\mathbb{R})$. The radial coordinate satisfies $r \geq \alpha_+$, so our analysis is restricted to the exterior of the black hole.\footnote{We do not consider the interior region here, although the orbifold identifications should be imposed on the full group manifold (see e.g.~\cite{Martinec:2002xq,Hemming:2002kd} for a discussion).}

The WZW model enjoys a chiral ${SL}(2,\mathbb R)_L \times { SL}(2,\mathbb R)_R$ symmetry. The associated current operators are given by 
\begin{equation}
J_{\rm sl}^a = \kk {\rm Tr}\left[ t^a_{\rm sl} \,\partial g_{\rm sl}\,g_{\rm sl}^{-1}\right]\,,\qquad \bar J_{\rm sl} ^a = \kk {\rm Tr}\left[t^a_{\rm sl}\,g_{\rm sl}^{-1}\,\bar\partial g_{\rm sl}\right]\,,
\end{equation}
where $t^a_{\rm sl}$ denote a convenient basis of ${SL}(2,\mathbb{R})$ generators (see Eq.~\eqref{eq:genSL}). The parametrization~\eqref{eq:BTZgroupmanifold} naturally selects a basis for the current algebra in which $J^3_0$ (the zero mode of $J^3$) is diagonal. Indeed, the generators of spacetime translations and rotations in this parametrization depend on $J^3_0$ and $\bar J_0^3$. 
This is known as the \emph{hyperbolic basis}. In contrast, the \emph{elliptic basis}, commonly used for global AdS$_3$~\cite{Maldacena:2000hw}, diagonalizes a different generator, which in our notation corresponds to $J_0^1$.   

In the elliptic basis, the ${ SL}(2,\mathbb{R})$ zero-mode algebra satisfies the commutation relations:
\begin{equation}
\left[ I_{\rm sl}^+ ,\,I_{\rm sl}^-\right] = -2 J^1_0\,,\qquad \left[ J^1_0,\,I_{\rm sl}^\pm\right] = \pm I_{\rm sl}^{\pm}\,,
\end{equation}
where $I^{\pm}_{\rm sl}= J_0^2 \pm i J_0^3$. In this case unitary irreducible representations are labelled by the real eigenvalue of $J_0^1$, denoted by $m$ and by the parameter $j$ that specifies the quadratic Casimir 
\begin{equation}
\label{eq:sl2casimir}
\mathbf C = \eta_{ab}J_0^a J_0^b=-j\left(j+1\right)\,.
\end{equation} 
As is well known, representations are classified into three main types:
\begin{itemize}
\item Principal continuous series $\mathcal C_j^{m_0}$. 

These representations contain states  $\ket{j; m_0, m}$, satisfying
\begin{equation}
J^1_0\ket{j; m_0, m} = m\ket{j; m_0, m}\,,
\end{equation}
with $m_0\in [0;1)$ and $m = m_0 + k$, $k\in \mathbb Z$. The spin parameter is taken as $j = 1/2 + i\nu$, $\nu<0$. 

\item Highest weight discrete series $\mathcal D_j^+$. 

These representations consist of states $\ket{j;m}$, with $m$ the eigenvalue of $J^1_0$,  $m = j-k$, $k\in\mathbb N$ and $j\leq -1/2$. The highest state $\ket{j;j}$ is annihilated by $I^+_{\rm sl}$, i.e. $I^+_{\rm sl} \ket{j;j}=0$. 

\item Lowest weight discrete series $\mathcal D_j^-$. 

Similarly, these representations are built from states $\ket{j;m}$, with  $m = -j+k$, $k\in\mathbb N$, and $j\leq -1/2$. The lowest state $\ket{j;-j}$ satisfies $I^-_{\rm sl} \ket{j;-j}=0$. 

\end{itemize}
Note that continuous representations have $\mathbf C \geq \frac{1}{4}$, while $\mathbf C \leq \frac{1}{4}$ for discrete series. 

In the hyperbolic basis, the relevant commutation relations are

\begin{equation} [J_0^+, J_0^-] = -2i J_0^3\,,\qquad [J_0^3, J_0^\pm] = \pm i J_0^\pm \,. \end{equation}

To label the states then, one diagonalizes $J_0^3$, whose eigenvalues $\lambda$ are continuous real numbers, unrelated to the Casimir parameter $j$. The representation space is built by acting with the negative modes of $J^a_n$ on primaries labelled by $(j,\lambda)$.

The Virasoro constraints~\cite{Natsuume:1996ij} in this setting read
\begin{equation}
\left( L_0 -1\right)= \left( -\frac{j\left(j+1\right)}{\kk -2} + N -1 + h_{\rm int}\right)\,,
\end{equation}
where $N$ is the worldsheet level number, and $h_{\rm int}$ accounts for the contribution from the internal manifold. A similar expression holds for the antiholomorphic sector.

\subsection{Spectral Flow}

The spectrum of string states on both global AdS$_3$ and BTZ includes a twisted sector, generated by spectral flow transformations. These operations generate additional, inequivalent representations that are essential for a complete description of the spectrum, as first emphasized in~\cite{Maldacena:2000hw} and later adapted to the BTZ case in~\cite{Hemming:2001we}. Spectral flow can be shown to be equivalent to the introduction of a twist operator that enforces specific identifications on the coordinates~\cite{Argurio:2000tb}. 
This analysis also imposes a quantization condition on the eigenvalues of $J^3_0$, which we will explicitly derive below.

We now discuss how spectral flow acts on string states in the BTZ geometry. Working within the parametrization introduced earlier, the transformation acts as
\begin{equation}
g \rightarrow \tilde g = {\rm e}^{\frac{1}{2}w_+ \,x^+\,\sigma_3}\,g\,{\rm e}^{-\frac{1}{2}w_- \,x^- \,\sigma_3}\,,
\end{equation}
where we denote worldsheet coordinates by $x^\pm=\tau_{\rm ws} \pm \sigma_{\rm ws}$. Under this transformation, the BTZ time and angular coordinates are mapped into
\begin{equation}
\begin{aligned}
\tau \rightarrow \tilde\tau = \tau + \frac{w_+ + w_-}{2}\tau_{\rm ws} + \frac{w_+ - w_-}{2}\sigma_{\rm ws}\,,\\[1mm]
\sigma \rightarrow \tilde\sigma = \sigma + \frac{w_+ - w_-}{2}\tau_{\rm ws} + \frac{w_+ + w_-}{2}\sigma_{\rm ws}\,.
\end{aligned}
\end{equation}
Imposing compatibility between the worldsheet periodicity $\sigma_{\rm ws} \sim \sigma_{\rm ws} + 2\pi$ and the global identifications defining the rotating BTZ geometry~\eqref{eq:btzglobalidentifications}, one finds that the spectral flow parameters must satisfy
\begin{equation}
\label{eq:spectralflowparameters}
w_\pm = \left( \alpha_+ \mp \alpha_-\right) \mathbf n\,,\qquad \mathbf n \in \mathbb Z\,.
\end{equation}
For a rotating BTZ black hole with $\alpha_\pm \neq 0$, this allows for asymmetric (i.e. independent) holomorphic and antiholomorphic spectral flows.

Spectral flow acts nontrivially on the current algebra and Virasoro generators. The modes of the current $J^3_{\rm sl}$ shift as
\begin{equation}
\begin{aligned}
J^3_n &\rightarrow \tilde J^3_n = J^3 _n + \frac{\kk}{2}w_+\,\delta_{n,0}\,,\qquad \bar J^3_n \rightarrow \tilde{\bar J}^3_n = \bar J^3_n - \frac{\kk}{2}w_-\,\delta_{n,0}\,,\\
\end{aligned}
\end{equation}
and the Virasoro generators are mapped to~\cite{Hemming:2001we}
\begin{equation}
L_n \rightarrow \tilde L_n = L_n + w_+ J^3_n + \frac{\kk}{4}w_+^2 \,\delta_{n,0}\,,\qquad \bar L_n \rightarrow \tilde{\bar L}_n = \bar L_n - w_- \bar J^3_n + \frac{\kk}{4}w_-^2 \,\delta_{n,0}\,.
\end{equation}

After spectral flow, the Virasoro constraints for physical states in the twisted sector become~\cite{Hemming:2001we,Rangamani:2007fz}
\begin{equation}
\label{eq:sl2Virasoro1}
- \frac{j\left( j +1 \right)}{\kk -2}-w_+\left(\lambda - \frac{\kk}{4}w_+\right)+ N + h_{\rm int} =1\,,
\end{equation}
and
\begin{equation}
\label{eq:sl2Virasoro2}
- \frac{j \left( j +1 \right)}{\kk -2}+w_-\left(\bar \lambda + \frac{\kk}{4}w_-\right)+ \bar N + \bar h_{\rm int} =1\,.
\end{equation} 
Here, $N$ and $\bar N$ denote oscillator levels and we have not considered states with fermion excitations for simplicity. 
Solving for $\lambda$ and $\bar \lambda$ gives,
\begin{equation}
\begin{aligned}
\lambda &= \frac{\kk}{4}w_+ - \frac{1}{w_+}\left( -\frac{ j\left(  j +1\right)}{\kk -2}+  N  -1 + h_{\rm int}\right)\,,\\[1mm]
\bar \lambda &= - \frac{\kk}{4}w_- + \frac{1}{w_-}\left(-\frac{ j\left( j +1\right)}{\kk -2}+  {\bar N} -1 + \bar h_{\rm int}  \right)\,.
\end{aligned}
\end{equation}
The level-matching condition then imposes a quantization condition:
\begin{equation}
\label{eq:quantcond}
\begin{aligned}
 N - \bar N &= \frac{\kk}{4}\left( w_-^2 - w_+^2\right) + w_- \bar\lambda + w_+  \lambda 
&= \mathbf n\,L\,,\qquad \mathbf n,\, L \in \mathbb Z\,,
\end{aligned}
\end{equation}
where $\mathbf n$ is the winding number around the non-contractible cycle generated by $\partial_\sigma$, and
\begin{equation}
L = \left( \alpha_+ - \alpha_-\right) \lambda + \left( \alpha_+ + \alpha_-\right) \bar \lambda + \mathbf n\,\kk\,\alpha_+ \,\alpha_- \in \mathbb Z\,,
\end{equation}
represents the quantized momentum carried by physical states.

These expressions resemble the corresponding results for global AdS$_3$ backgrounds (see e.g.~\cite{Maldacena:2000hw}), but with important distinctions. 
The first major difference concerns the allowed values of the spectral flow parameters. Formally, the AdS$_3$ case can be recovered from~\eqref{eq:spectralflowparameters} by setting $\alpha_+ =0$ and $\alpha_- =-1$, leading to $w_+ = -w_- \equiv w \in \mathbb Z$. Thus, in AdS$_3$, the spectral flow is symmetric between the holomorphic and antiholomorphic sectors.
Second, the structure of the spectrum is different. In AdS$_3$, the eigenvalue $m$ of $J_0^1$ in the flowed sector is
\begin{equation}
\label{eq:mforAdS}
m = \frac{\kk}{4}w + \frac{1}{w}\left(-\frac{ j\left(  j +1\right)}{\kk -2}+  N  -1 + h_{\rm int} \right) \,,
\end{equation}
and an analogous expression holds for $\bar m$. Consider the spectral flow of a primary state in the highest weight discrete series. For a highest-weight discrete series representation, $m = j - q$ for some $q \in \mathbb N$. Solving~\eqref{eq:mforAdS} for $j$ shows that only discrete spin values appear, implying a discrete energy spectrum for these spectrally flowed representations -- these are the so-called \emph{short strings} confined within AdS$_3$. By contrast, continuous representations (\emph{long strings}) do not impose such constraints and yield a continuous energy spectrum.

Now, returning to the BTZ background, regardless of the initial unflowed representation, there is no fixed relation between the $J_0^3$ eigenvalue $\lambda$ and $j$. As a result, this difference implies that all states in the twisted sector exhibit a continuous energy spectrum.


\section{Absence of CTCs for supersymmetric solutions}
\label{sec:CTC}

To analyze the conditions under which supersymmetric solutions are free of closed timelike curves (CTCs) both on and outside the horizon, we adapt the argument of~\cite{Chong:2005hr} to our six-dimensional setup, which can be viewed as an uplift of the case considered there. Supersymmetric solutions in six dimensions always admit a Killing vector $V$, constructed as a bilinear of Killing spinors, which is everywhere null~\cite{Gutowski:2003rg}. In our conventions this vector is $V = \partial_t + \partial_y$. To isolate the null coordinate, we introduce 
\begin{equation}
u = t\,,\qquad v = y-t\,,
\end{equation}
so that the metric~\eqref{eq:susysolution} can be rewritten as
\begin{equation}
\begin{aligned}
{\rm d }s^2 &= -\Delta_u {\rm d}u^2 + Q_5\left(A_r \,{\rm d}\hat r^2 + {\rm d}\theta^2\right) + B_v \left( {\rm d}v + v_0 \,{\rm d}u\right)^2\\[1mm]
&\quad + B_\phi \left( {\rm d}\phi + v_1\, {\rm d}v+ v_2\,{\rm d}u\right)^2 + B_\psi \left( {\rm d}\psi + v_3\,{\rm d}v + v_4 \,{\rm d}u\right)^2\,,
\end{aligned}
\end{equation}
with the functions $\Delta_u$, $B_v$, $B_\phi$, $B_\psi$, $v_{0,1,2,3,4}$ determined by comparison with~\eqref{eq:susysolution}. The null condition on $V= \partial_u$ implies that its norm satisfies
\begin{equation}
\label{eq:nullnorm}
-\Delta_u + B_v \,v_0^2 + B_\phi \, v_2^2 + B_\psi \, v_4^2=0\,.
\end{equation}
Since 
\begin{equation}
\lim_{\hat r\rightarrow 0} \Delta_u = 0\,,
\end{equation}
it follows that at the Killing horizon one necessarily has $B_v \cdot B_\phi \cdot B_\psi < 0$, which signals the presence of naked CTCs. There are, however, two special cases in which CTCs can be avoided, giving rise instead to supersymmetric extremal black holes or to smooth topological solitons.

It is also useful to examine the Euclidean section of these solutions, where one can formally associate a supersymmetric inverse temperature~\cite{Anupam:2023yns},
\begin{equation}
\label{eq:susy_beta}
    \beta = \pi Q_5\left( Q_1 + Q_p\right) \left( \frac{1}{\sqrt{4 Q_1 Q_5 Q_p - J_-^2}}- \frac{i}{J_+}\right)\,,
\end{equation}
which is complex, as is typical of supersymmetric non-extremal geometries.

\paragraph{The BMPV black hole.}

The first way to avoid CTCs is to impose
\begin{equation}
\lim_{\hat r\rightarrow 0}v_0 = \lim_{\hat r\rightarrow 0}v_2 =\lim_{\hat r\rightarrow 0}v_4= 0\,,
\end{equation}
so that the null condition~\eqref{eq:nullnorm} is automatically satisfied. This occurs in the extremal limit $\beta \to \infty$,
\begin{equation}
    b\to 0\,\implies\, J_+ \to 0\,.
\end{equation}
In this regime, the functions $B_{v,\phi,\psi}$ behave as
\begin{equation}
\lim_{\hat r\rightarrow 0} B_v = \frac{Q_p}{Q_1}\left(1-\ell^2\right)\,,\qquad \lim_{\hat r\rightarrow 0} B_\phi = Q_5 \,\sin^2\theta\,,\qquad \lim_{\hat r\rightarrow 0} B_\psi =  Q_5\,\cos^2\theta\,.
\end{equation}
Thus naked CTCs are absent provided
\begin{equation}
Q_{1,5,p}>0,\,\qquad 1-\ell^2 >0 \quad  \Rightarrow \quad 4 Q_1 Q_5 Q_p - J_-^2 >0\,.
\end{equation}
This is precisely the regime studied in section~\ref{sec:susyblackhole}.

In this case, the Killing horizon at $\hat r = 0$ corresponds to the event horizon of a well-defined supersymmetric black hole that is smooth on and outside the horizon, with finite, real entropy~\cite{Breckenridge:1996is}.

\paragraph{The topological soliton.} The second way to avoid CTCs is to require
\begin{equation}
\lim_{\hat r\rightarrow 0}\Delta_u \neq 0\,,\qquad \lim_{\hat r\rightarrow 0}B_v = 0\,.
\end{equation}
This is realized by taking
\begin{equation}
\label{eq:solitonlimit}
b\rightarrow 4\frac{\sqrt{\ell^2-1}\,Q_1\,Q_p}{Q_1 + Q_p}\,,
\end{equation}
which remains real only for $\ell > 1$. This corresponds precisely to the supersymmetric analogue of~\eqref{eq:decsoliton}, yielding a smooth horizonless solution. The resulting topological soliton is the supersymmetric version of the horizonless geometries studied in~\cite{Giusto:2004kj,Jejjala:2005yu,Martinec:2018nco} (see e.g.~\cite{Giusto:2004id,Giusto:2004ip,Giusto:2012yz}), where a spacelike circle contracts in the interior (possibly leading to conical singularities) and the Lorentzian geometry caps off without a horizon. 

In the limit~\eqref{eq:solitonlimit}, the angular momenta of~\eqref{eq:bhcharges} reduce to
\begin{equation}
J_- = 2\ell\,\sqrt{Q_1 Q_5 Q_p}\,, \qquad J_+ =2\sqrt{Q_1Q_5Q_p\left(\ell^2-1\right)}\,.
\end{equation}
Requiring the solution to remain Lorentzian, i.e. that the charges are real, enforces
\begin{equation}
\ell^2 - 1 >0 \quad \Rightarrow \quad 4 Q_1 Q_5 Q_p - J_-^2 <0\,.
\end{equation}
In this solitonic regime the charges then satisfy the constraint 
\begin{equation}
    J_+^2 = J_-^2 - 4 Q_1 Q_5 Q_p\,,
\end{equation}
which implies that the supersymmetric inverse temperature~\eqref{eq:susy_beta} vanishes in the limit \eqref{eq:solitonlimit},
\begin{equation}
    b\rightarrow 4\frac{\sqrt{\ell^2-1}\,Q_1\,Q_p}{Q_1 + Q_p}\,\implies \,\beta \to 0\,.
\end{equation}


\bibliography{ws_bh.bib}
\bibliographystyle{JHEP}

\end{document}